\documentclass[reprint,superscriptaddress,amsmath,amssymb,aps,prb,onecolumn,12pt]{revtex4-2}

\usepackage{graphicx}
\usepackage{dcolumn}
\usepackage{bm}
\usepackage{tabularx}
\usepackage{hyperref}
\usepackage{float}

\begin{document}

\title{Near- and Far-Field Excitation of Topological Plasmonic Metasurfaces}

\author{Matthew Proctor}
    \email{matthew.proctor12@imperial.ac.uk}
    \affiliation{Department of Mathematics, Imperial College London, London SW7 2AZ, UK}

\author{Xiaofei Xiao}
    \affiliation{Blackett Laboratory, Department of Physics, Imperial College London, London SW7 2AZ, United Kingdom}
    
\author{Richard V. Craster}
    \affiliation{Department of Mathematics, Imperial College London, London SW7 2AZ, UK}
    
\author{Stefan A. Maier}
    \affiliation{Blackett Laboratory, Department of Physics, Imperial College London, London SW7 2AZ, United Kingdom}
    \affiliation{Chair in Hybrid Nanosystems, Nano-Institute Munich, Faculty of Physics, Ludwig-Maximilians-Universit\"at M\"unchen, 80539 M\"unchen, Germany}
 
\author{Vincenzo Giannini}
    \affiliation{Instituto de Estructura de la Materia (IEM), Consejo Superior de Investigaciones Científicas (CSIC), Serrano 121, 28006 Madrid, Spain}
 
\author{Paloma Arroyo Huidobro}
    \affiliation{Instituto de Telecomunica\c c\~oes, Insituto Superior Tecnico-University of Lisbon, Avenida Rovisco Pais 1, 1049-001 Lisboa, Portugal}

\date{\today}

\begin{abstract}
The breathing honeycomb lattice hosts a topologically non-trivial bulk phase due to the crystalline-symmetry of the system. Pseudospin-dependent edge states which emerge at the interface between trivial and non-trivial regions can be used for directional propagation of energy. Using the plasmonic metasurface as an example system, we probe these states in the near and far-field using a semi-analytical model. 
We give the conditions under which directionality is observed and show that it is source position dependent.
By probing with circularly-polarised magnetic dipoles out of the plane, we first characterize modes along the interface in terms of the enhancement of source emission due to the metasurface. We then excite from the far-field with non-zero orbital angular momentum beams. 
The position dependent directionality holds true for all classical wave systems with a breathing honeycomb lattice. Our results show that a metasurface in combination with a chiral two-dimensional material could be used to guide light effectively on the nanoscale.
\end{abstract}

\maketitle

\section{Introduction}

Topological nanophotonics offers a path towards efficient and robust control over light on the nanoscale \cite{rider2019perspective}. Concepts borrowed from topological insulators, materials which host protected surface states for electrons, can also be applied to photonic systems. Following theoretical proposals \cite{haldane2008possible,raghu2008analogs}, protected photonic modes reliant on a explicit time-reversal symmetry breaking component have been demonstrated experimentally \cite{wang2009observation}. However, these require strong magnetic fields or complex materials with a large magneto-optical response, which makes such systems difficult to miniaturise. More recently these effects have been proposed using graphene \cite{jin2017infrared, pan2017topologically}, which naturally has a large magneto-optical response. This is limited to the infrared regime however, which restricts potential applications. Methods of achieving topological protection through the crystalline symmetry of a system whilst preserving time-reversal symmetry are therefore appealing, since they do not require complex setups and are not restricted to a specific frequency regime. These effects fall into two main categories: valley effects, which rely on extrema in the band structures of materials \cite{makwana2018geometrically, wong2020gapless, proctor2020manipulating}, and pseudospin-dependent effects, which rely on the spin angular momentum texture of the electromagnetic fields  \cite{saba2020nature, orazbayev2019quantitative}. 

First proposed in Ref. \cite{wu2015scheme}, the pseudospin-dependent effect relies on a triangular lattice of hexamers to produce states reminiscent of the quantum spin Hall effect (QSHE) in topological insulators. This breathing honeycomb lattice has two gapped phases: the shrunken phase, where the hexamers are perturbed inwards and the expanded phase where they are perturbed outwards. Despite both having a trivial $\mathbb{Z}_2$ index, the phases of the breathing honeycomb are topologically distinct: while the shrunken phase is a trivial insulator, the expanded one is an instance of an `obstructed atomic limit' phase \cite{depaz2019engineering,proctor2020robustness}, and edge states will appear between regions in either phase. A direct analogy of the QSHE would produce purely unidirectional edge states, in the absence of spin-mixing impurities. However, as we showed in a previous work \cite{proctor2019exciting}, the edge mode directionality for near-field sources is source-position dependent and is determined by the spin angular momentum of the modes: it is the local handedness of the elliptical field polarisation which determines the propagation direction of the edge modes, rather than an absolute pseudospin quantity as in the QSHE \cite{proctor2019exciting}. This result is true for any bosonic breathing honeycomb lattice \cite{oh2018chiral} since it is rooted in time-reversal symmetry and the absence of Kramer's degeneracy for bosons as opposed to fermions. There have been a range of experimental investigations on the breathing honeycomb photonic crystal \cite{smirnova2019third, barik2016two, barik2018topological, yves2017crystalline}, including the observation of edge modes in the visible regime \cite{parappurath2020topological, Liu2020photonic}. Despite this, there is no comprehensive theoretical study of the methods for exciting pseudospin edge modes and in particular, the necessary conditions for exciting unidirectional modes. 

In this article, we consider a plasmonic metasurface consisting of a two-dimensional array of metallic nanoparticles with a breathing honeycomb lattice. We first characterise the optical response of the bulk modes and then investigate the propagation properties of edge states which emerge at the interface between trivial and and non-trivial regions. We then extend the understanding of excitation by near-field sources by probing edge states with sources out of the plane. Finally, we show that the propagation of edge states with far-field beams is determined by the electric field phase of the edge eigenmodes . Whilst the results we present are in the plasmonic metasurface, we emphasise that the properties and behaviour of the edge states are valid for any classical wave system.

\section{Methods}

We model the system of subwavelength, metallic nanoparticles (NPs) using the coupled dipole method. When the nearest neighbour spacing $R$ and NP radius $r$ satisfy $R > 3r$, each NP can be treated as a point dipole \cite{maier2007plasmonics, weber2004propagation}. To model nanorods, we use spheroidal NPs with radius $r=10$~nm and height $h=40$~nm. For NPs of this size and shape it is necessary to include depolarization and radiative effects and so we use the Meier Wokaun long wavelength approximation (MWLWA) to describe the the dipolar optical response of an individual NP  \cite{moroz2009depolarization, meier1983depolarization}. The MWLWA polarisability $\alpha(\omega)$ for spheroids is,
\begin{align}
    \alpha(\omega) = \frac{\alpha_s(\omega)}{1 - \frac{k^2}{l_E}D\alpha_s(\omega) - i\frac{2k^3}{3}\alpha_s(\omega)},
\end{align}
with the static polarisability $\alpha_s(\omega)$,
\begin{align}
    \alpha_s(\omega) = \frac{V}{4\pi}\frac{\epsilon(\omega) - 1}{1 + L(\epsilon(\omega) - 1)}.
\end{align}
$k = \epsilon_m \omega/c$ is the wave number and the environment is a homogeneous vacuum with $\epsilon_m = 1$. $V$ is the NP volume and $D$ and $L$ are dynamic and static geometrical factors, and $l_E$ is the spheroid half-axis; $D=1$, $L=1/3$ and $l_E = r$ for a sphere. The dielectric function $\epsilon(\omega)$ is given by the Drude model,
\begin{align}
\epsilon(\omega) = \epsilon_\infty - \frac{\omega_p^2}{\omega^2 + i\omega\gamma}.
\end{align}
We use silver NPs, with $\epsilon_\infty = 5$, $\omega_p = 8.9$~eV and $\gamma = 1/17$~fs $\approx38$~meV \cite{yang2015optical}. 
For a system of multiple NPs, we can write a self-consistent coupled dipole equation which describes the dipole moment of a NP due to neighbouring NPs as well as an incident electric field $\mathbf{E}_\mathrm{inc}$,
\begin{align}
    \frac{1}{\alpha(\omega)}\mathbf{p}_i = \mathbf{E}_\mathrm{inc} + \hat{\mathbf{G}}(\mathbf{d}_{ij}, \omega)\cdot\mathbf{p}_j.
\end{align}
The interactions between dipoles are characterised by the dyadic Green's function \cite{abajo2007colloquium},
\begin{align}\label{eqn:dyadic_gf}
    \hat{\textbf{G}}(\textbf{d}_{ij}, \omega) &= k^2\frac{e^{ikd}}{d} \biggl[
    \biggl(
    1 + \frac{i}{kd} - \frac{1}{k^2d^2}
    \biggr)\hat{\textbf{I}} \,- 
    \biggl(
    1 + \frac{3i}{kd} - \frac{3}{k^2d^2}
    \biggr)\textbf{n}\otimes\textbf{n}
    \biggr],
\end{align}
where $\mathbf{d}_{ij}$ is the separation between dipole $i$ and $j$, $d = |\mathbf{d}_{ij}|$ and $\mathbf{n} = \mathbf{d}_{ij}/d$. (The $\hat{\cdot}$ represents a dyadic operator.) We only take the $zz$-component of the Green's function, which corresponds to interactions between dipole moments perpendicular to the separation between the NPs (i.e. out-of-plane). This is a valid approximation since we use spheroidal NPs, causing the in-plane modes to become well separated in frequency from the out-of-plane modes \cite{wang2016existence}. 
For a periodic array of NPs in a plasmonic metasurface we can apply Bloch's theorem and write the following system of equations,
\begin{align}\label{eqn:matrix_problem}
    \left(\hat{\mathbf{I}}\frac{1}{\alpha(\omega)} - \hat{\textbf{H}}(\textbf{k}_\parallel, \omega)\right)\cdot\textbf{p} = \mathbf{E}_\mathrm{inc}. 
\end{align}
The vector $\mathbf{p}$ contains the out-of-plane dipole moments $p_z$ of all NPs in the unit cell. The interaction matrix $\hat{\textbf{H}}(\textbf{k}_\parallel, \omega)$ has elements,
\begin{align}\label{eqn:interaction_matrix}
    H_{pq}=
    \begin{cases}
    \sum\limits_{\textbf{R}} \hat{\textbf{G}}(\textbf{d}_{pq} + \textbf{R}, \omega) \hspace{2px} e^{i\textbf{k}_\parallel\cdot\textbf{R}} & p \neq q\\
    \sum\limits_{|\textbf{R}|\neq0} \hat{\textbf{G}}(\textbf{R}, \omega) \hspace{2px} e^{i\textbf{k}_\parallel\cdot\textbf{R}} & p = q
    \end{cases}\hspace{1em},
\end{align}
where $p$, $q$ are unit cell indices. The summations run over the lattice sites $\mathbf{R}=n\mathbf{a}_1 + m\mathbf{a}_2$, with lattice vectors $\mathbf{a}_1$ and $\mathbf{a}_2$. The interaction matrix has dimension $N\times N$ where $N$ is the number of NPs in the unit cell. Since the lattice sums are slowly converging, we use Ewald's method to calculate them \cite{linton2010lattice,Kolkowski2020lattice}. When modelling plasmonic metasurfaces, it is necessary to include the long range, retarded interactions in the Green's function in \autoref{eqn:dyadic_gf}. Whilst the quasistatic approximation can be used to model very subwavelength plasmonic chains and arrays \cite{koenderink2006complex, zhen2008collective}, it fails to accurately capture the behaviour of modes at the light line and the radiative broadening and redshifting of modes, which becomes apparent in NPs with $r > 10$~nm. Additionally, retarded interactions can affect the topological properties of some plasmonic systems \cite{pocock2018topological, pocock2019bulk}. 

By solving \autoref{eqn:matrix_problem}, we can calculate the extinction cross section $\sigma_\mathrm{ext}$ from the dipole moments and incident electric field using the optical theorem,
\begin{align}
    \sigma_\mathrm{ext}= \frac{k\,\mathrm{Im}(\mathbf{p} \cdot \mathbf{E}^*_\mathrm{inc})}{|\mathbf{E}^2|}.
\end{align}
The incident field satisfies Maxwell's equations, $|\mathbf{k}_\parallel| E_\parallel + k_z E_z = 0$ (we assume $E_\parallel = 1$ and harmonic time dependence $e^{-i\omega t}$). When the field is propagating, above the light line, $k_z = \sqrt{k^2 - |\mathbf{k}_\parallel|^2}$ and when it is evanescent, below the light line, $k_z = i\sqrt{|\mathbf{k}_\parallel|^2 - k^2}$. We can also solve \autoref{eqn:matrix_problem} as an eigenvalue problem and calculate the spectral function, by letting $\mathbf{E}_\mathrm{inc} = 0$, which allows us to probe modes whether they are bright or dark. The spectral function $\sigma_\mathrm{sf}$ is,
\begin{align}
    \sigma_\mathrm{sf} = k\,\mathrm{Im}(\alpha_\mathrm{eff}),
\end{align}
with the effective polarisability $\alpha_\mathrm{eff} = \sum_i 1/\lambda^{(i)}$ for eigenvalues $\lambda^{(i)}$ \cite{zhen2008collective}. Finally, the coupled dipole method can also be applied to finite systems to model electromagnetic scattering \cite{Merchiers2007Light}. 


\section{Optical Response of Bulk and Edge States}\label{sec:optical_response}

\begin{figure}[t]
\centering
\includegraphics[width=\linewidth]{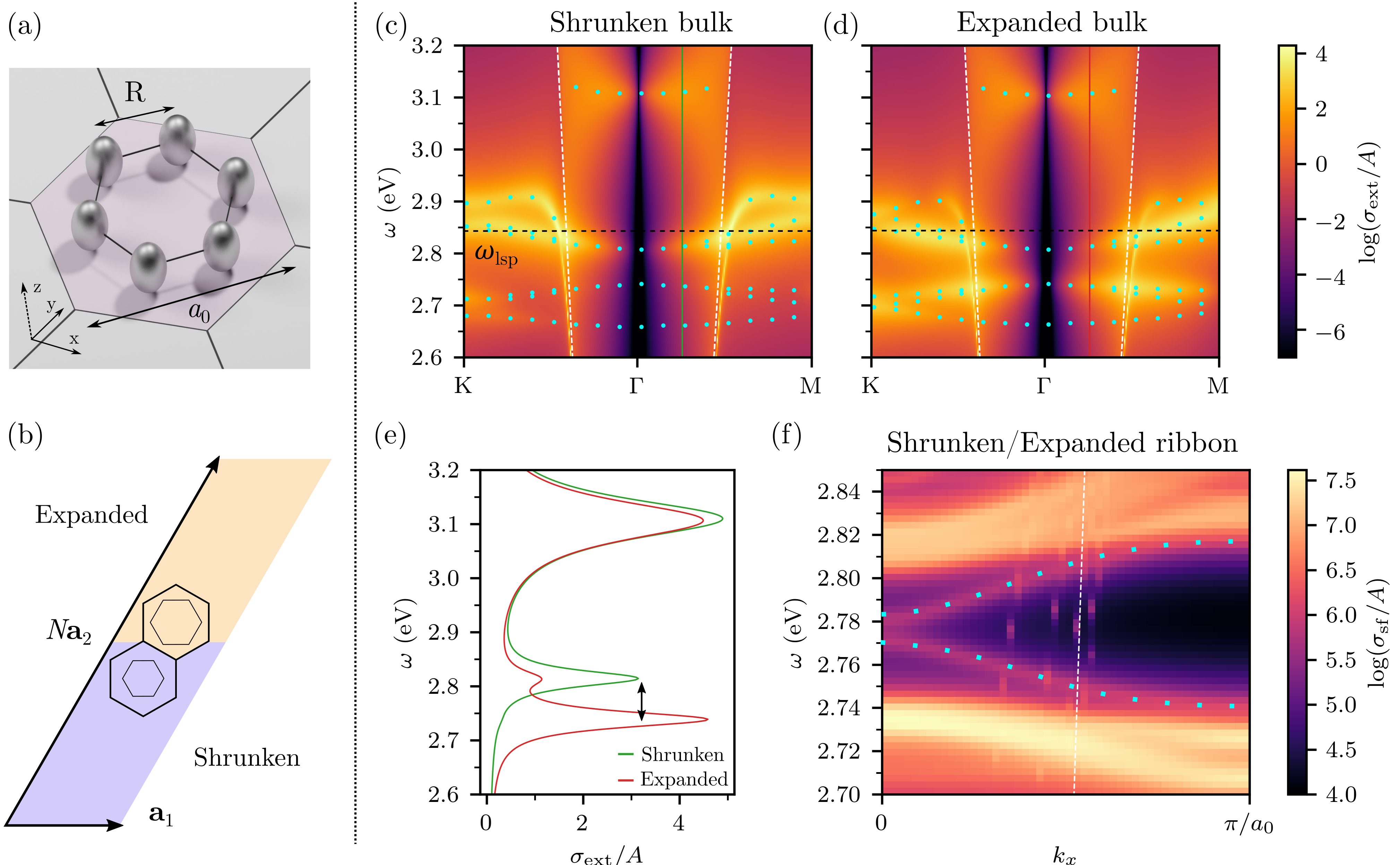}
\caption{Lattice set up and optical response: (\textbf{a}) Unit cell with lattice vectors $\mathbf{a}_1 = (a_0,0)$ and $\mathbf{a}_2 = (a_0/2, \sqrt{3}/2a_0)$, the lattice constant $a_0 = 120$~nm. The honeycomb lattice has nearest neighbour spacing $R = R_0 = a_0/3$. The expanded lattice has $R = 1.065 R_0$ and the shrunken lattice $R=0.9 R_0$.
(\textbf{b}) Ribbon interface between expanded (orange) and shrunken (blue) regions. The ribbon is doubly periodic, with lattice vectors $\mathbf{a}_1$ and $N\mathbf{a}_2$. The total ribbon length is $N=20$, with $10$ expanded and $10$ contracted unit cells.
Extinction cross sections $\sigma_\mathrm{ext}$ for (\textbf{c}) shrunken and (\textbf{d}) expanded lattices. The band structures, found from the spectral function, are highlighted in blue. (\textbf{e}) $\sigma_\mathrm{ext}$ for the shrunken and expanded lattices at $\mathbf{k}_\parallel$ shown in (\textbf{c, d}). A band inversion occurs near $\Gamma$, causing the dipolar band to switch with the quadrupolar band (arrow).
(\textbf{f}) Spectral function $\sigma_\mathrm{sf}$ for the ribbon interface in (\textbf{e}). Drude losses are $\gamma = 10$~meV to increase the visibility of the edge states and they are highlighted (blue dots).
}
\label{fig:bulk_edge_response}
\end{figure}

We begin by characterising the optical response of the periodic system, in order to later excite the finite system at the correct frequencies. The breathing honeycomb lattice setup is shown in \autoref{fig:bulk_edge_response}(a) with a unit cell with six NPs \cite{wu2015scheme}. In a honeycomb lattice, $R = R_0 = a_0/3$ where $R$ is the nearest neighbour spacing and $a_0$ is the lattice constant. We let $R_0 = 40$~nm, so the lattice constant $a_0 = 120~$nm. The larger unit cell (compared to the rhombic Wigner-Seitz unit cell) folds the Brillouin zone and results in a double Dirac cone at $\Gamma$ (as shown in Appendix \ref{sec:appendix_honeycomb}, \autoref{fig:appendix_honeycomb}). We note that the Dirac cone lies below the localised surface plasmon frequency $\omega_{lsp}$ due to chiral symmetry breaking, long range interactions \cite{proctor2019exciting}. By perturbing the nearest neighbour separation with a scale factor $s$, such that $R = sR_0$, a band gap is opened at $\Gamma$. In \autoref{fig:bulk_edge_response}(c, d) we plot the extinction cross section $\sigma_\mathrm{ext}$ across the Brillouin zone for the shrunken ($R = 0.9R_0$) and expanded ($R=1.065R_0$) lattices. The scale factors are chosen to ensure that the size of the band gaps for the two lattices approximately equal. We plot the bands from the spectral function as blue dots on top of the extinction cross section as a guide. Below the light line (white dotted line), where modes are confined, the modes have high quality factors whereas above the light line the modes are very broad corresponding to larger radiative losses.

In the shrunken lattice, for $\mathbf{k}_\parallel$ close to $\Gamma$, the NPs in the unit cell hybridise to form hexapoles, quadrupoles, dipoles and monopoles (from lowest energy to highest energy). The monopolar mode has dipole moments in phase, forming a bonding mode in the out-of-plane dipole moments across the whole lattice. As a result it couples very strongly with the light line and exhibits a polariton-like splitting. We note that the ordering in this plasmonic metasurface is opposite to the photonic crystal due to the metallic nature of the NPs \cite{proctor2019exciting}.
A band inversion occurs around $\Gamma$ between the shrunken and expanded phases, causing the dipolar and quadrupolar bands to flip. This is similar to the band inversion process in the QSHE.
Although the bands become much broader above the light line, there is still a clear signature between shrunken and expanded phases which is observable in the far-field, as has been shown previously \cite{proctor2019exciting, gorlach2018far}. We plot $\sigma_\mathrm{ext}$ for a fixed wavevector $\mathbf{k}_\parallel$ in  \autoref{fig:bulk_edge_response}(e), showing how the dipolar peak shifts between the two phases. The broad peak at the top corresponding to the monopolar mode does not shift. Unlike the QSHE, neither the shrunken or expanded lattices have a non-trivial $\mathbb{Z}_2$-invariant. However, they are still topologically distinct and one way to distinguish them is by their Wilson loops \cite{depaz2020tutorial}. The shrunken lattice is topologically trivial whilst the expanded lattice is in an photonic `obstructed atomic limit' phase \cite{depaz2019engineering}; it is the $C_6$ symmetry which provides the topological character.

Despite having trivial $\mathbb{Z}_2$-invariants, when regions in the shrunken phase are placed next to regions in the expanded phase, edge states will appear the band gap due to the topological origin of the band inversion \cite{proctor2019exciting}. These edge states are not topologically protected but do have a pseudospin character, which allows directional modes to be excited through a similar mechanism to the chiral light matter interactions in photonic crystals \cite{lodahl2017chiral}. We model the edge states of the system by setting up a ribbon with an interface between the two phases, as shown in \autoref{fig:bulk_edge_response}(b). The ribbon supercell is $N=20$ unit cells along the $\mathbf{a}_2$ direction, with 10 unit cells in the expanded and shrunken phases, respectively. In \autoref{fig:bulk_edge_response}(f), we plot the spectral function $\sigma_\mathrm{sf}$ for the ribbon for $\mathbf{k}_\parallel = 0$ to $\pi/a_0$. (The spectral function from $\mathbf{k}_\parallel = 0$ to $-\pi/a_0$ is identical.) Edge states appear in the band gap from $\omega \approx 2.74$~eV to $2.81~$eV and a small minigap appears between $\omega \approx 2.77$~eV to $2.78$~eV. The mini-gap appears due to the interface breaking $C_6$ symmetry, which is the symmetry that protects the topological phase. We note that we use an `armchair' interface here but it is also possible to define other terminations for this lattice \cite{Kariyado2017}, including the `zig-zag' interface \cite{orazbayev2019quantitative}. In the latter case, the $C_6$ symmetry breaking across the interface is slightly smaller and results in a smaller mini-gap, but otherwise the behaviour of the edge states is qualitatively similar. This is evident in the hybridisation of NPs close to the interface:
the lower band has an anti-bonding character whilst the upper band has a bonding character (we show these in Appendix \ref{sec:appendix_edge_hybridisation}, \autoref{fig:appendix_edge_hybridisation}). Due to doubly-periodic supercell, we see multiple narrow bands bending down and crossing the band gap in \autoref{fig:bulk_edge_response}(f). This is due to diffraction orders above the light line (white dotted line) \cite{cherqui2019plasmonic} and does not affect the investigation of edge states which follows. 

\section{Circularly-Polarized Point Sources}\label{sec:point_sources}

We begin by extending the investigation of our previous work \cite{proctor2019exciting} by exciting the system with point sources with circularly-polarized magnetic fields, $H = H_x \pm i H_y$. 
The directionality of modes is well understood for sources in the plane of the metasurface, $h=0$: The inhomogeneous spin angular momentum of the edge eigenmode determines the direction of propagation, rather than the polarization of the source. This means it is possible to selectively couple to either of the counter-propagating edge states by moving the source in the plane of the metasurface; a result which holds for any bosonic system with this lattice, including photonic crystals \cite{oh2018chiral} (We show this explicitly in Appendix \ref{sec:photonic_crystal}).
It is important to understand whether it is still possible to excite edge modes for sources placed out of the plane, including whether directionality is still observed. This is relevant to experimental setups where quantum dots and emitters, or 2D materials are placed on top of a spacing layer above the NPs in the metasurface. 

Experimentally, it is challenging to realise circularly-polarized magnetic dipoles at optical frequencies and in nanoscale setups, which would be required to excite directional modes in our metasurface by near-field sources. Previous experiments showed the coupling of Zeeman-induced circularly-polarized excited states of quantum dots to the directional edge modes in a photonic crystal \cite{barik2018topological}. However, the directionality in our metasurface is related to the circular polarisation of the in-plane magnetic field, meaning a magnetic rather than electric point source is required. 
At low frequencies these sources have been realized by means of coaxial cables \cite{yves2017crystalline,yves2020locally}. 
On the other hand, magnetic transitions in the optical regime can be realized with rare earth ions \cite{baranov2017modifying}, although these are usually linearly-polarized transitions. A recent proposal makes use of two anti-parallel atomic dipoles to generate a magnetic dipole at optical frequencies \cite{alaee2020quantum}, which could be extended to circular polarization, and possibly realized with quantum dots. Alternatively, the magnetic resonance of split-ring resonators could be engineered to provide the required source. 
In the following, we assume a simple model and use a dielectric NP as source, given that they support a magnetic dipole mode \cite{garcia2011strong} which could be excited with a circularly polarised wave.
Provided the system can be modelled in the dipole approximation with subwavelength resonators and point dipole sources, the following results are valid regardless of the physical nature of the source or frequency regime. Therefore our main conclusions also apply to metasurfaces made of microwave \cite{yves2017crystalline,yves2020locally} or Mie resonators \cite{minkyung2020quantum}.

\begin{figure}[t]
\centering
\includegraphics[width=\linewidth]{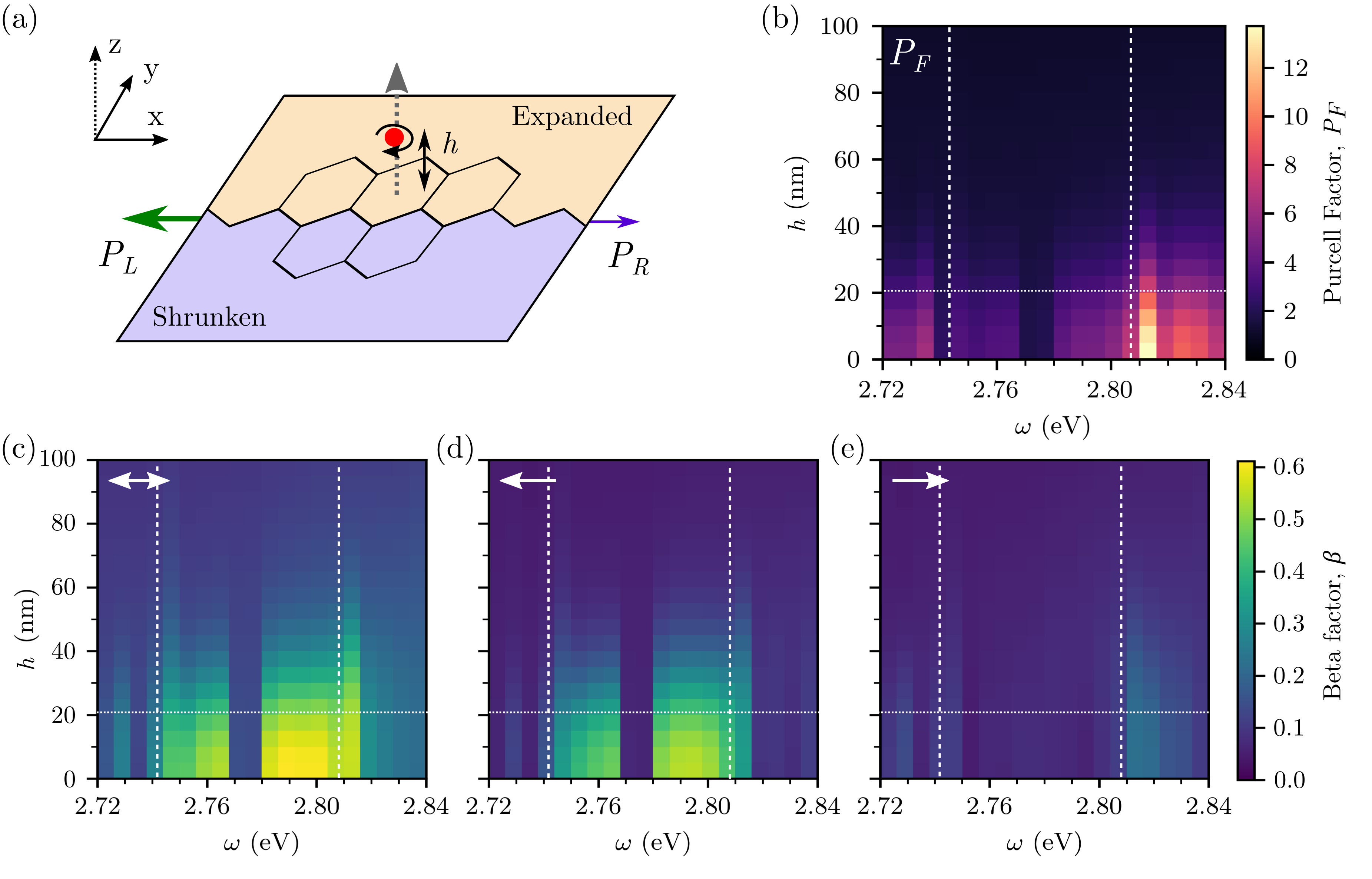}
\caption{Single sources at varying height $h$ and frequency $\omega$: (\textbf{a}) Schematic showing the layout of the interface. The system is 40 unit cells along $x$ and 12 unit cells along $y$, with 6 contracted and 6 expanded unit cells. A left circularly-polarized source (red) is placed at the centre of an expanded unit cell and moved upwards. The power through the left $P_L$ and right $P_R$ channels is used to characterise the directionality. 
(\textbf{b}) Purcell factor $P_F$ of the source. The height of the NPs is shown as a horizontal dotted line and the edges of the band gap are shown as vertical lines. Beta factors: (\textbf{c}) Power coupling into the edge $\beta_\mathrm{edge}$ (double arrow). Power coupling into the (\textbf{d}) left $\beta_{L}$, and (\textbf{e}) right $\beta_{R}$ channels (left and right arrows).}
\label{fig:single_source_out_plane}
\end{figure}

The scattering setup is shown in \autoref{fig:single_source_out_plane}(a). 
The source position is moved to various heights $h$ in the $z$-direction and the $xy$-position is fixed at the centre of the expanded unit cell, as shown in \autoref{fig:single_source_out_plane}(a). The power through the left ($P_L$) and right ($P_R$) channels is calculated, as well as the power radiated by the source. The material losses along the interface are set to zero to test the directionality behaviour and they are gradually increased at the edge of the sample to prevent backscattering. 

We define a $\beta$-factor to characterise the amount of power coupling into edge modes compared to the total power radiated by the source,
\begin{align}
    \beta_i = \frac{P_i}{P_0 P_F},
\end{align}
where $i = (L, R)$ corresponding to power through the left or right channel, respectively. $P_0$ is the power radiated by the source and $P_F$ is the Purcell factor. The total power coupling into the edge is then $\beta_\mathrm{edge} = \beta_\mathrm{L} + \beta_\mathrm{R}$. If $\beta_\mathrm{edge} = 0$, none of the energy radiated from the source couples into the edge mode, whereas if $\beta_\mathrm{edge} = 1$ all of the energy couples into the edge. $\beta$ accounts for enhancement of the emission of the source due to the environment through the Purcell factor \cite{baranov2017modifying}. For a magnetic dipole the emitted power in free space,
\begin{align}
    P_0 = \mu_0\frac{\omega^4}{3c^3}|\mathbf{m}|^2,
\end{align}
with vacuum permeability $\mu_0$ and magnetic dipole moment $\mathbf{m}$.

The source is placed
initially at $h=0$~nm and then moved upwards to $h=100$~nm. We use a left circularly-polarized dipole throughout ($H = H_x - iH_y$); which, in combination with the $xy$-position at the centre of the expanded unit cell, means we expect it to couple to the left propagating edge mode. Additionally, we scan over the frequency range of the band gap to investigate whether exciting in the upper or lower band has an effect on the directionality. We measure the normalised power through $yz$-planes perpendicular to both the metasurface and the interface. First, in \autoref{fig:single_source_out_plane}(b) we plot the Purcell factor. (A horizontal white dashed line serves as a guide to the height of the NPs in the metasurface and the band gap is highlighted with vertical dashed lines). We see that the Purcell factor is greatest near the metasurface since the array of NPs enhances the emission of the source by increasing the available local density of states (LDOS). For $h > 50~$nm, there is almost no coupling to the metasurface and $P_F \approx 1$. Scanning across frequencies at $h$ close to zero, we can clearly see the minigap near $2.77$~eV, where no NPs are excited which causes $P_F$ to decrease. Lastly, $P_F$ is largest at $\omega \approx 2.81$~eV where the source begins to excite bulk modes.

Next, in panel (c) we plot $\beta_\mathrm{edge}$ where it is clear that it is greatest for the edge state frequencies and smallest in the minigap and for bulk frequencies; in agreement with the spectral function of the edge states in \autoref{fig:bulk_edge_response}(f). A maximum of approximately $60\%$ of the power emitted by the source couples into the metasurface at $h=0$, and the rest radiates into free space as the source emits in all directions and the bulk of the metasurface is gapped. As the source is moved upwards, $\beta_\mathrm{edge}$ decreases and follows the same relationship as $P_F$ such that for $h > 50~$nm very little energy couples into the edge state. This confirms that the increase of $P_F$ is due to the source coupling to the edge. Finally, in \autoref{fig:single_source_out_plane} (d, e) we plot the beta factors for the left $\beta_L$ and right $\beta_R$ edge channels. As explained previously, since we use a left circularly-polarized source which is initially placed at the centre of the unit cell, we expect and observe power flow predominantly through the left channel \cite{proctor2019exciting}, i.e. $\beta_\mathrm{L} \gg \beta_\mathrm{R}$. This pattern of directionality holds as the source is moved upwards. 


\begin{figure}[t]
\centering
\includegraphics[width=\linewidth]{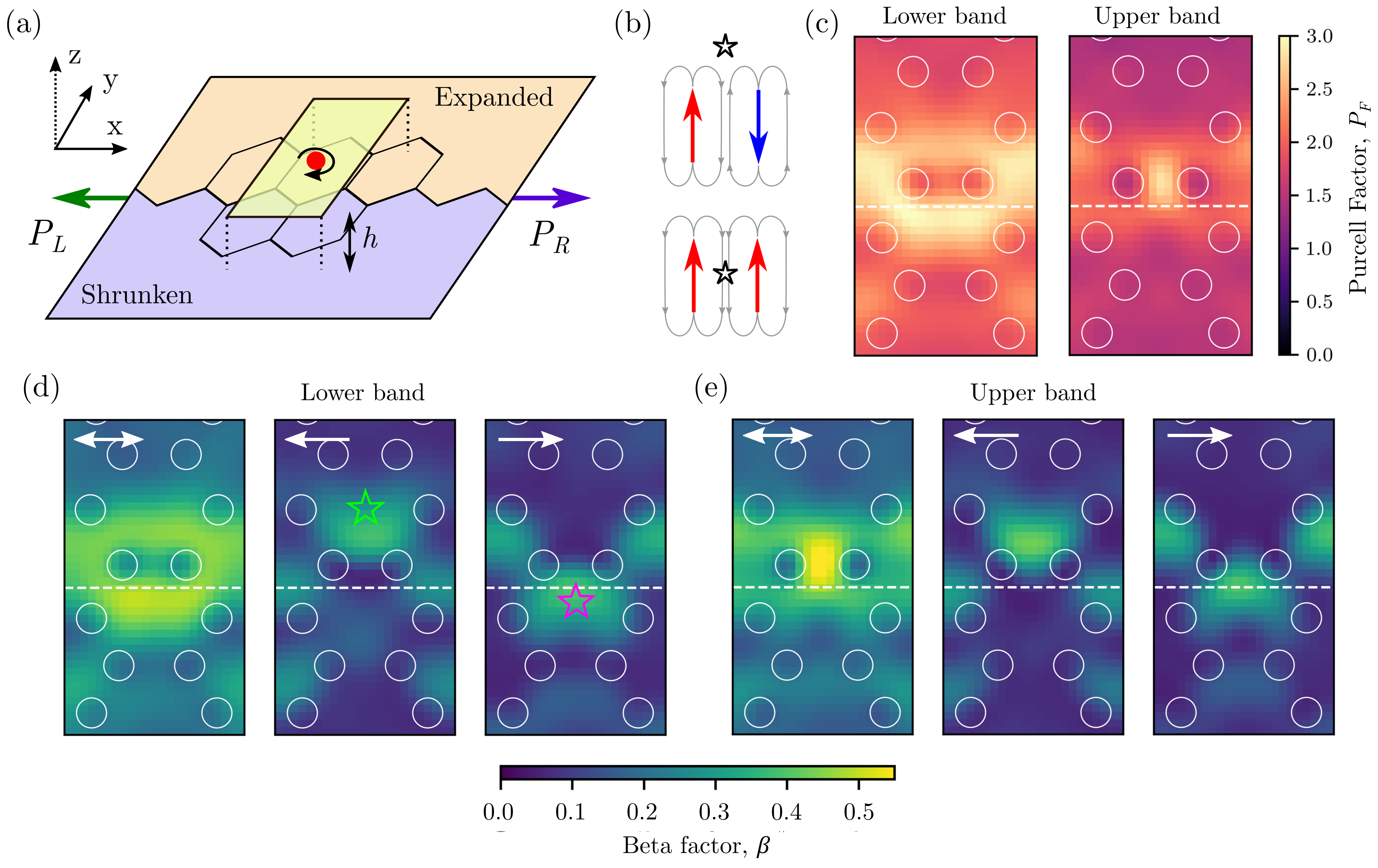}
\caption{Single sources for varying $xy$-position across the interface and fixed height $h=30$~nm. (\textbf{a}) Schematic showing the interface and source layout. (\textbf{b}) Hybridisation of out-of-plane dipole moments in NPs for the the bonding upper band $\omega = 2.795$~eV and anti-bonding lower band $\omega=2.76$~eV. The maximum electric field is at the star in each case. (\textbf{c}) Purcell factors, for lower (left) and upper (right) bands. Beta factors for the  (\textbf{d}) lower and (\textbf{e}) upper band. Left panels $\beta_\mathrm{edge}$, power coupling into the edge, middle panels $\beta_{L}$, power coupling into the left and right panels $\beta_{R}$, power coupling into the right channels.}
\label{fig:single_source_in_plane}
\end{figure}   

Following the understanding of directionality for sources at different heights, we now investigate the position dependence in a plane parallel to the metasurface.
The sources are placed in a range of positions in the $xy-$plane at a fixed height, $h=30$~nm, close to the interface as shown in \autoref{fig:single_source_in_plane}(a). We choose this height to ensure a $10~$nm gap between the top of the NPs and the excitation layer, and to ensure enough power couples into the edge. In an experimental setup, this gap will be dependent on the type of source used. We will excite the system in the upper band at $\omega = 2.795~$eV and in the lower band at $\omega = 2.76~$eV.

In \autoref{fig:single_source_in_plane}(c) we plot the Purcell factor for the lower (left) and upper (right) bands: The pattern of each band is very different. From the ribbon eigenmodes in \autoref{fig:bulk_edge_response}(f) (and Appendix \ref{sec:appendix_edge_hybridisation}, \autoref{fig:appendix_edge_hybridisation}), we showed that the upper and lower bands have a bonding and anti-bonding character. In \autoref{fig:single_source_in_plane}(b) we sketch the out-of-plane dipole moments for the two NPs closest to the edge. This shows how the electric field maxima and minima will be in different positions: For the anti-bonding mode there is a maximum (star) at some position above the NPs and for the bonding mode, the maximum is between the NPs. This is reflected in the peaks of the Purcell factor. Furthermore, for both bands the Purcell factor is greatest at the interface which again is due to the LDOS being greatest here. Away from the interface, it approaches unity but there is still some emission enhancement due to the proximity to the metasurface. However, this means that sources placed far from the interface will not excite edge modes.

Finally, in \autoref{fig:single_source_in_plane}(d, e) we plot the beta factors for the chosen frequencies in the lower and upper bands. $\beta_\mathrm{edge}$ follows the same pattern as the Purcell factor. However, it is interesting to note that despite the upper and lower bands having different $\beta_\mathrm{edge}$ patterns, the position dependent directionality can still be seen in $\beta_L$ (middle panels) and $\beta_R$ (right panels). Specifically, for both bands the coupling to the left propagating channel for a left circularly-polarized source is maximum when the source is above the centre of an expanded unit cell (lime star). 
As soon as the source is moved outside the hexamer of NPs (magenta star), it couples to the mode travelling in the opposite direction. For larger $h$ this pattern will eventually be lost as the source couples less and less to the edge state.

\section{Far-field Circularly-polarized Excitations}\label{sec:far_field}
A number of experimental works have investigated the directionality of these edge states under a far-field excitation, in a range of frequency regimes \cite{smirnova2019third, parappurath2020topological, Liu2020photonic}. Despite this, there is no detailed theoretical investigation of directionality of edge modes with a far-field excitation and specifically whether the position-dependent directionality found for point sources also holds here. 
We investigate this by mimicking a circularly-polarized far-field Gaussian beam excitation with non-zero orbital angular momentum (OAM) incident on the metasurface. 
The NPs are excited with an incident electric field with a Gaussian intensity profile, $\mathbf{E}_\mathrm{inc} = (0, 0, E_z)$ with,
\begin{align}\label{eqn:beam}
    E_z \propto e^{-r^2/(2w^2)} e^{i\phi},
\end{align}
$r$ is centre of the beam, $w$ is the full width at half maximum (FWHM), $\phi$ is the angle in radians from the centre. The phase vortex in $E_z$ corresponds to a left or right circular polarization.

\begin{figure}[t]
    \centering
    \includegraphics[width=\linewidth]{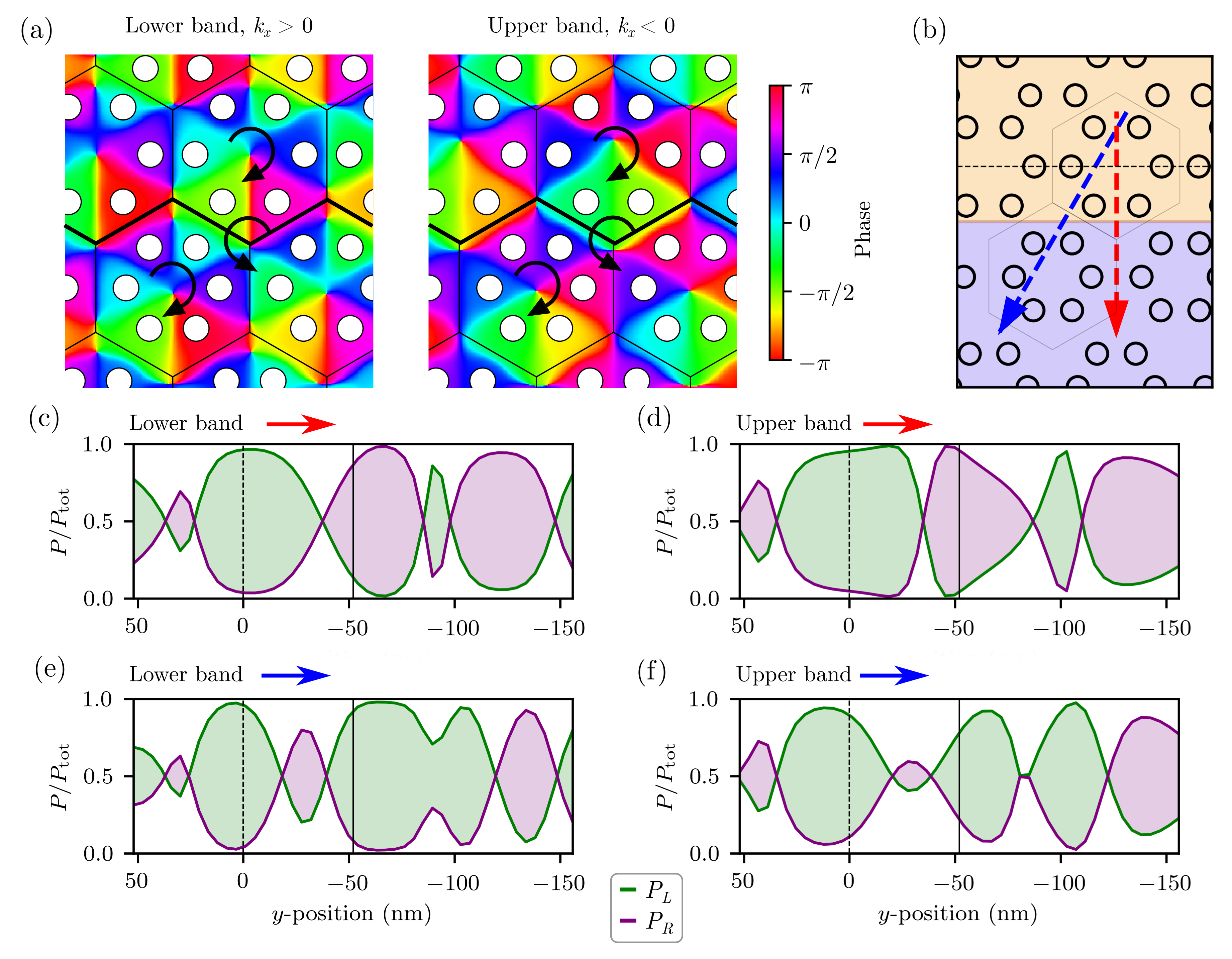}
    \caption{Far field excitations: 
    (\textbf{a}) Phase of the $z$-component of the electric field for edge eigenmodes, in the (left) lower band for $k_x > 0$ and (right) upper badn for $k_x < 0$. An incident beam with non-zero orbital angular momentum will couple to the phase vortices (black arrows) of the edge modes. 
    (\textbf{b}) A diagram showing the paths of the beam across the interface between expanded (orange) and shrunken (blue) regions. The $y$-value of the centre of the expanded unit cell is shown as a dotted black line.
    (\textbf{c}, \textbf{d}) Beams along the red path. Fraction of power to the left $P_L/P_\mathrm{tot}$ (green) and right $P_R/P_\mathrm{tot}$ (purple) channels upper band $\omega=2.795$~eV and the lower band $\omega=2.76$~eV. Regions are coloured to indicate directionality to the left or right. $y$-values for the centre of the unit cell (vertical dotted line) and interface (vertical solid line) are highlighted. 
    (\textbf{e}, \textbf{f}) Beam along the blue path. Fraction of power to the left and right for lower and upper bands. This path is chosen to maximise directionality.}
    \label{fig:beam_directionality}
\end{figure}

In \autoref{fig:beam_directionality}(a), we plot the phase of the $z$-component of the electric field of the edge eigenmodes: for the lower band, at $k_x > 0$ and for the upper band, at $k_x < 0$. A phase vortex structure is present across the interface, with clearly distinct regions at the centres of unit cells and at the edge of unit cells.
Mode matching and maximum directionality occurs when the phase vortex of the beam rotates in the the same direction as the phase vortex of the edge eigenmode \cite{deng2017transverse}, these positions are highlighted as clockwise arrows. Mode mismatch occurs when the beam and eigenmode have vortices rotating in opposite directions and these positions are shown are anti-clockwise arrows. When there is a complete mode mismatch, the beam will excite a mode travelling in the opposite direction.
The phase vortex behaviour of directional edge states has previously been studied in the context of valley modes \cite{deng2019vortex, chen2017valley, chen2018tunable, ye2017observation}, where valleys at $K$ and $K'$ have vortices rotating in opposite directions. 

To confirm this behaviour we will probe the finite system with the beam in \autoref{eqn:beam}. We let $w = 50~$nm, meaning the beam covers approximately one unit cell and the centre of the beam is moved over the red and blue paths shown in \autoref{fig:beam_directionality}(b). As in the previous section, we excite the system at frequencies in the lower and upper bands at $\omega=2.76$~eV and $\omega=2.795$~eV. In panels (c) and (d) we plot the fraction of power through the left ($P_L/P_\mathrm{tot}$) and right ($P_R/P_\mathrm{tot}$) channels for the two bands (plots of the total power coupling to the edge modes in each case are shown in Appendix \ref{sec:appendix_power}). Importantly, when the centre of the beam is at the centre of the expanded unit cell ($y=0$~nm), the majority of power is through the left channel $P_L \gg P_R$, as was the case with point sources. Similarly, when the beam moves across the interface ($y\approx-50~$nm) the majority of power switches to the right channel $P_L \ll P_R$. Although we have only demonstrated this for two specific frequencies here, the position dependent directionality holds for frequencies across the band gap (as we show in Appendix \ref{sec:appendix_power}, \autoref{fig:appendix_power}). 
The vortex map in \autoref{fig:beam_directionality}(a) shows that it is possible to choose a path which maximises directionality, by avoiding traversing over phase singularities.
In \autoref{fig:beam_directionality}(d) we plot the power for the blue path shown in (a). This path is similar to scanning perpendicularly across a zig-zag interface, as in \cite{parappurath2020topological}. Compared to the red path, there is wider range of $y$-values that yield the expected directionality. Finally, we emphasise that it is the relationship between the FWHM and the lattice constant, rather than the FWHM value itself, which is important. For beams with a FWHM which is much larger than the lattice constant, directionality is mostly lost since the beam expands to cover and excite the majority of the interface. 

In an experimental setup, edge modes in the metasurface could be excited with a combination of a 2D layer and far-field beam. For example, valley-selective modes can be excited using a transition metal dichalcogenide on top of plasmonic metasurface \cite{chervy2018room,hu2019coherent,sun2019separation}. Along with the results from \autoref{sec:point_sources}, the directionality observed for far-field excitations suggests that a similar method could be employed to excite directional modes in the breathing honeycomb lattice. 


\section{Conclusions}

In this article we have provided a comprehensive study of the excitation of edge modes in a plasmonic metasurface with a breathing honeycomb lattice arrangement. The 2D lattice of metallic NPs hosts subwavelength pseudospin edge modes which arise due to the topology of the bulk. The plasmonic metasurface is a versatile system for testing the directionality of these modes through the coupled dipole method. Motivated by the excitation of chiral and valley selective modes in plasmonic metasurfaces with 2D layers, we probe the edge states of our system in the near- and far-field. With circularly-polarised magnetic dipole sources, we map the directionality of modes for sources out of the plane and show how, provided the source still couples to the interface, the pattern of directionality predicted by spin angular momentum is preserved. Additionally, we probe edge modes with far-field beams with non-zero orbital angular momentum. Here, the direction of propagation is predicted by the phase of the $E_z$ field of the edge eigenmodes. Importantly, although we have particularized to the plasmonic metasurface, the directionality behaviour of the edge modes is applicable to any classical wave system possessing the same breathing honeycomb lattice.

\begin{acknowledgments}
M.P. and P.A.H. acknowledge funding from the Leverhulme Trust. P.A.H. acknowledge financial support from Funda\c c\~ao para a Ci\^encia e a Tecnologia and Instituto de Telecomunica\c c\~oes under projects UID/EEA/50008/2020 and the CEEC Individual program with reference CEECIND/03866/2017.
\end{acknowledgments}

\clearpage

\appendix

\section{Honeycomb lattice: Optical response of bulk modes}\label{sec:appendix_honeycomb}

\begin{figure}[H]
    \centering
    \includegraphics[width=\linewidth]{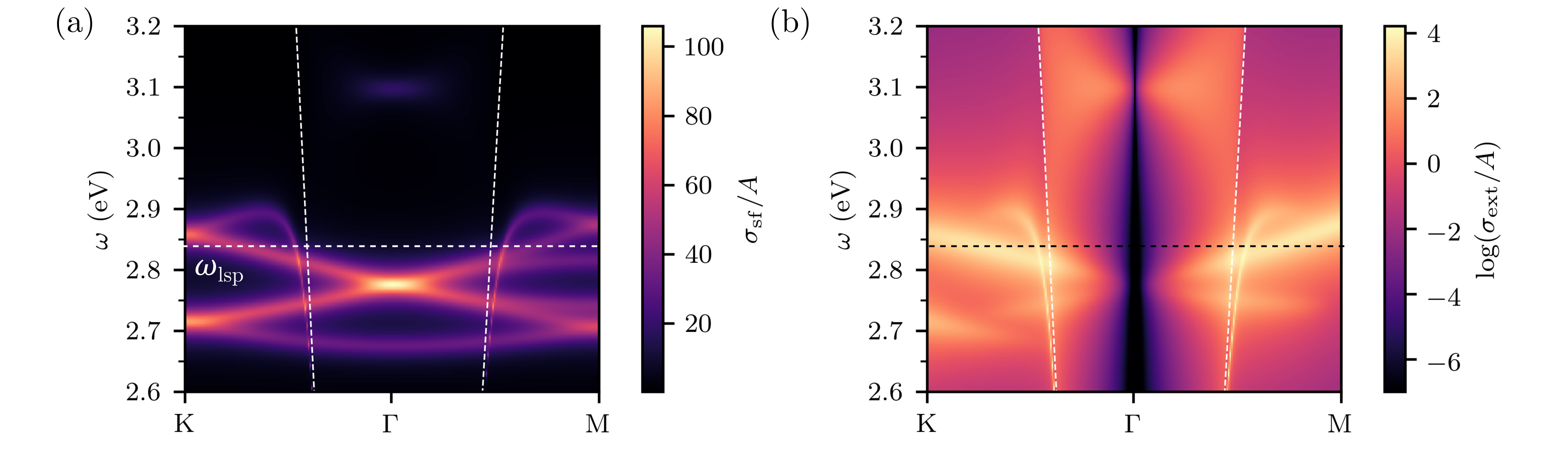}
    \caption{(\textbf{a}) Spectral function $\sigma_\mathrm{ext}$. There is a double Dirac cone degeneracy at $\Gamma$ due to the folded Brillouin zone, (\textbf{b}) Extinction cross section $\sigma_\mathrm{ext}$.}
    \label{fig:appendix_honeycomb}
\end{figure} 

\section{Edge state eigenmodes}\label{sec:appendix_edge_hybridisation}

\begin{figure}[H]
    \centering
    \includegraphics[width=\linewidth]{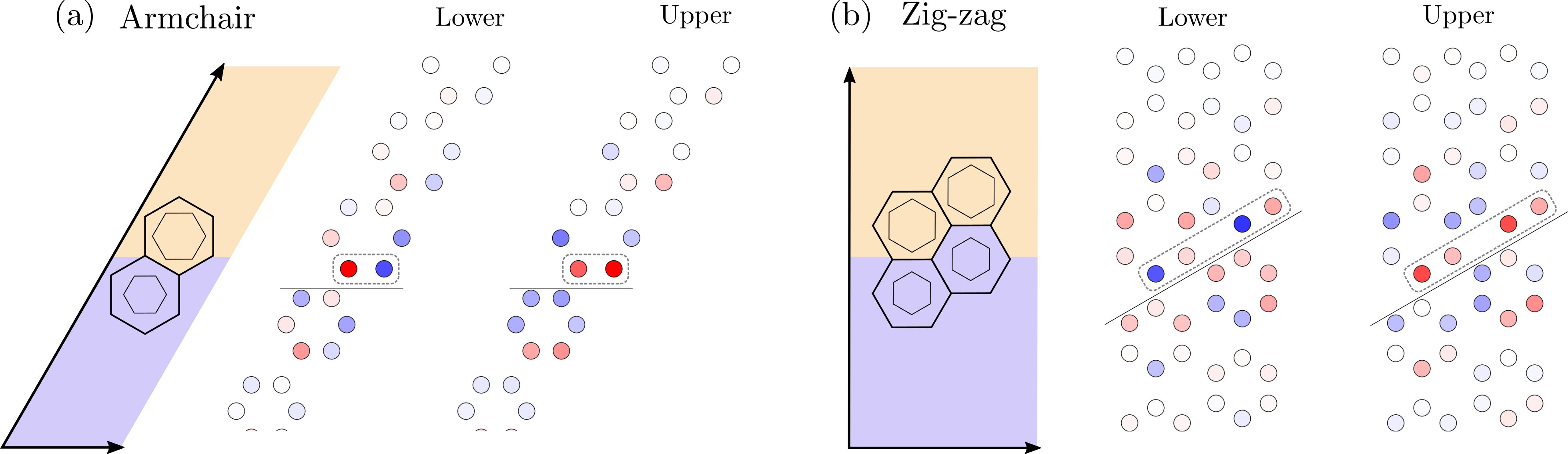}
    \caption{Hybridisation of NPs at interface: (\textbf{a}) Armchair interface, (\textbf{b}) Zig-zag interface}
    \label{fig:appendix_edge_hybridisation}
\end{figure}

The system of equations in \autoref{eqn:matrix_problem} for the ribbon interface is solved as an eigenvalue problem by letting $\mathbf{E}_\mathrm{inc} = 0$. Additionally, the Green's function is linearised by letting $\omega = \omega_{\mathrm{lsp}}$, the localized surface plasmon frequency, in order to calculate dipole moments $\mathbf{p}$. The dipole moments for the armchair and zig-zag interfaces are shown in \autoref{fig:appendix_edge_hybridisation}.

\section{Directionality of modes in a photonic crystal}\label{sec:photonic_crystal}

In the main text, we particularise to the subwavelength, plasmonic metasurface. We will show that the near-field position dependent directionality also holds in a photonic crystal, with constant permittivity, dielectric elements. 
We perform finite element simulations in COMSOL \cite{comsol} on the same breathing honeycomb interface. The contracted region has $s = 0.967$ and the expanded $s = 1.041$, and the dielectric pillars are silicon with $\epsilon = 11.7$ \cite{wu2015scheme}. 

\begin{figure}[H]
    \centering
    \includegraphics[width=\textwidth]{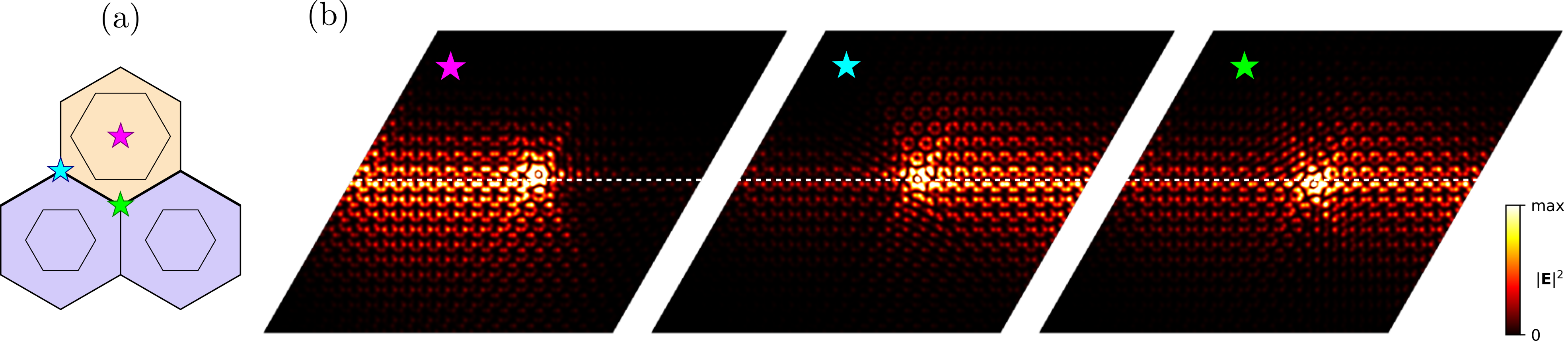}
    \caption{Breathing honeycomb photonic crystal interface. (a) Left circularly-polarised sources are placed at three positions near the interface. (b) Electric field intensity $|\mathbf{E}|^2$ for the different source positions, demonstrating source position dependent directionality.}
    \label{fig:photonic_crystal}
\end{figure}

A left circularly-polarized magnetic dipole is used to excite modes at the three positions shown in \autoref{fig:photonic_crystal}(a). The electric field intensity $|\mathbf{E}|^2$ of the interface for the three source positions is shown in (b).
When the source is at the centre of the expanded unit cell (magenta star), it couples to a left propagating mode. Whereas when the source is directly at the interface (green and cyan stars), it couples predominantly to a right propagating mode.

\section{Far field excitations: Power through left and right channels}\label{sec:appendix_power}

Power through the left and right channels is calculated in the same way as in \autoref{sec:far_field}, however here we plot the total power through the edge (normalised to the maximum) rather than the fraction left/right. As the beam moves downwards, we see how peaks in $P_\mathrm{tot}$ follow a qualitatively similar pattern to the beta factors calculated in \autoref{fig:single_source_in_plane}(b, d), which reflects the incident beam coupling to the edge modes. 
A maximum in $P_\mathrm{tot}$ for the lower band occurs as the centre of the beam passes through the centre of two particles immediately at the interface, whereas for the upper band $P_\mathrm{tot}$ peaks either side of this. 

\begin{figure}[H]
    \centering
    \includegraphics[width=\linewidth]{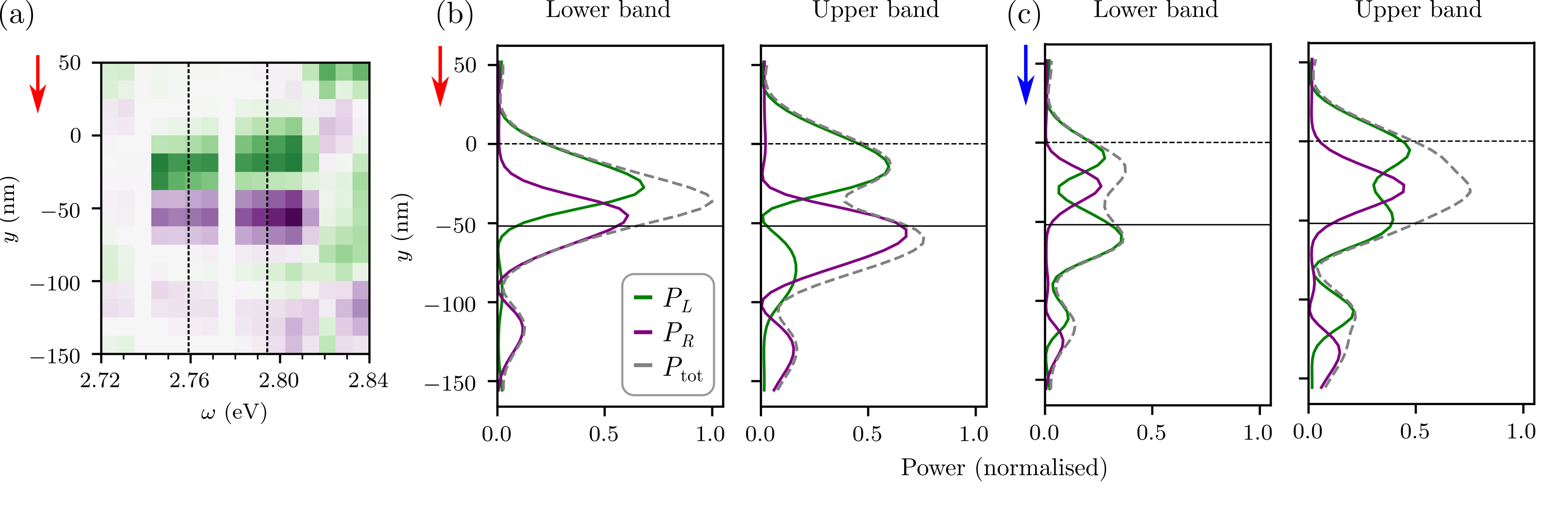}
    \caption{Far field excitations: 
    (\textbf{a}) $P_\Delta = P_L - P_R$ for the red path in \autoref{fig:beam_directionality}(b) and frequencies across the band gap. 
    (\textbf{b}) Directionality along the red path: Power through the edge $P_\mathrm{tot}$, left $P_{L}$ and right channels $P_{R}$ for lower and upper bands. (The lower and upper band line plots are equivalent to the vertical lines in (\textbf{a}).)
    (\textbf{c}) Directionality along the blue path.}
    \label{fig:appendix_power}
\end{figure}   

\clearpage
\bibliography{main}

\begin{thebibliography}{55}%
\makeatletter
\providecommand \@ifxundefined [1]{%
 \@ifx{#1\undefined}
}%
\providecommand \@ifnum [1]{%
 \ifnum #1\expandafter \@firstoftwo
 \else \expandafter \@secondoftwo
 \fi
}%
\providecommand \@ifx [1]{%
 \ifx #1\expandafter \@firstoftwo
 \else \expandafter \@secondoftwo
 \fi
}%
\providecommand \natexlab [1]{#1}%
\providecommand \enquote  [1]{``#1''}%
\providecommand \bibnamefont  [1]{#1}%
\providecommand \bibfnamefont [1]{#1}%
\providecommand \citenamefont [1]{#1}%
\providecommand \href@noop [0]{\@secondoftwo}%
\providecommand \href [0]{\begingroup \@sanitize@url \@href}%
\providecommand \@href[1]{\@@startlink{#1}\@@href}%
\providecommand \@@href[1]{\endgroup#1\@@endlink}%
\providecommand \@sanitize@url [0]{\catcode `\\12\catcode `\$12\catcode
  `\&12\catcode `\#12\catcode `\^12\catcode `\_12\catcode `\%12\relax}%
\providecommand \@@startlink[1]{}%
\providecommand \@@endlink[0]{}%
\providecommand \url  [0]{\begingroup\@sanitize@url \@url }%
\providecommand \@url [1]{\endgroup\@href {#1}{\urlprefix }}%
\providecommand \urlprefix  [0]{URL }%
\providecommand \Eprint [0]{\href }%
\providecommand \doibase [0]{https://doi.org/}%
\providecommand \selectlanguage [0]{\@gobble}%
\providecommand \bibinfo  [0]{\@secondoftwo}%
\providecommand \bibfield  [0]{\@secondoftwo}%
\providecommand \translation [1]{[#1]}%
\providecommand \BibitemOpen [0]{}%
\providecommand \bibitemStop [0]{}%
\providecommand \bibitemNoStop [0]{.\EOS\space}%
\providecommand \EOS [0]{\spacefactor3000\relax}%
\providecommand \BibitemShut  [1]{\csname bibitem#1\endcsname}%
\let\auto@bib@innerbib\@empty
\bibitem [{\citenamefont {Rider}\ \emph {et~al.}(2019)\citenamefont {Rider},
  \citenamefont {Palmer}, \citenamefont {Pocock}, \citenamefont {Xiao},
  \citenamefont {Arroyo~Huidobro},\ and\ \citenamefont
  {Giannini}}]{rider2019perspective}%
  \BibitemOpen
  \bibfield  {author} {\bibinfo {author} {\bibfnamefont {M.~S.}\ \bibnamefont
  {Rider}}, \bibinfo {author} {\bibfnamefont {S.~J.}\ \bibnamefont {Palmer}},
  \bibinfo {author} {\bibfnamefont {S.~R.}\ \bibnamefont {Pocock}}, \bibinfo
  {author} {\bibfnamefont {X.}~\bibnamefont {Xiao}}, \bibinfo {author}
  {\bibfnamefont {P.}~\bibnamefont {Arroyo~Huidobro}},\ and\ \bibinfo {author}
  {\bibfnamefont {V.}~\bibnamefont {Giannini}},\ }\bibfield  {title} {\bibinfo
  {title} {A perspective on topological nanophotonics: Current status and
  future challenges},\ }\href {https://doi.org/10.1063/1.5086433} {\bibfield
  {journal} {\bibinfo  {journal} {Journal of Applied Physics}\ }\textbf
  {\bibinfo {volume} {125}},\ \bibinfo {pages} {120901} (\bibinfo {year}
  {2019})}\BibitemShut {NoStop}%
\bibitem [{\citenamefont {Haldane}\ and\ \citenamefont
  {Raghu}(2008)}]{haldane2008possible}%
  \BibitemOpen
  \bibfield  {author} {\bibinfo {author} {\bibfnamefont {F.~D.~M.}\
  \bibnamefont {Haldane}}\ and\ \bibinfo {author} {\bibfnamefont
  {S.}~\bibnamefont {Raghu}},\ }\bibfield  {title} {\bibinfo {title} {Possible
  realization of directional optical waveguides in photonic crystals with
  broken time-reversal symmetry},\ }\href
  {https://doi.org/10.1103/PhysRevLett.100.013904} {\bibfield  {journal}
  {\bibinfo  {journal} {Phys. Rev. Lett.}\ }\textbf {\bibinfo {volume} {100}},\
  \bibinfo {pages} {013904} (\bibinfo {year} {2008})}\BibitemShut {NoStop}%
\bibitem [{\citenamefont {Raghu}\ and\ \citenamefont
  {Haldane}(2008)}]{raghu2008analogs}%
  \BibitemOpen
  \bibfield  {author} {\bibinfo {author} {\bibfnamefont {S.}~\bibnamefont
  {Raghu}}\ and\ \bibinfo {author} {\bibfnamefont {F.~D.~M.}\ \bibnamefont
  {Haldane}},\ }\bibfield  {title} {\bibinfo {title} {Analogs of
  quantum-hall-effect edge states in photonic crystals},\ }\href
  {https://doi.org/10.1103/PhysRevA.78.033834} {\bibfield  {journal} {\bibinfo
  {journal} {Phys. Rev. A}\ }\textbf {\bibinfo {volume} {78}},\ \bibinfo
  {pages} {033834} (\bibinfo {year} {2008})}\BibitemShut {NoStop}%
\bibitem [{\citenamefont {Wang}\ \emph {et~al.}(2009)\citenamefont {Wang},
  \citenamefont {Chong}, \citenamefont {Joannopoulos},\ and\ \citenamefont
  {Solja{\v{c}}i{\'{c}}}}]{wang2009observation}%
  \BibitemOpen
  \bibfield  {author} {\bibinfo {author} {\bibfnamefont {Z.}~\bibnamefont
  {Wang}}, \bibinfo {author} {\bibfnamefont {Y.}~\bibnamefont {Chong}},
  \bibinfo {author} {\bibfnamefont {J.~D.}\ \bibnamefont {Joannopoulos}},\ and\
  \bibinfo {author} {\bibfnamefont {M.}~\bibnamefont {Solja{\v{c}}i{\'{c}}}},\
  }\bibfield  {title} {\bibinfo {title} {Observation of unidirectional
  backscattering-immune topological electromagnetic states},\ }\href
  {https://doi.org/10.1038/nature08293} {\bibfield  {journal} {\bibinfo
  {journal} {Nature}\ }\textbf {\bibinfo {volume} {461}},\ \bibinfo {pages}
  {772} (\bibinfo {year} {2009})}\BibitemShut {NoStop}%
\bibitem [{\citenamefont {Jin}\ \emph {et~al.}(2017)\citenamefont {Jin},
  \citenamefont {Christensen}, \citenamefont {Solja\ifmmode \check{c}\else
  \v{c}\fi{}i\ifmmode~\acute{c}\else \'{c}\fi{}}, \citenamefont {Fang},
  \citenamefont {Lu},\ and\ \citenamefont {Zhang}}]{jin2017infrared}%
  \BibitemOpen
  \bibfield  {author} {\bibinfo {author} {\bibfnamefont {D.}~\bibnamefont
  {Jin}}, \bibinfo {author} {\bibfnamefont {T.}~\bibnamefont {Christensen}},
  \bibinfo {author} {\bibfnamefont {M.}~\bibnamefont {Solja\ifmmode
  \check{c}\else \v{c}\fi{}i\ifmmode~\acute{c}\else \'{c}\fi{}}}, \bibinfo
  {author} {\bibfnamefont {N.~X.}\ \bibnamefont {Fang}}, \bibinfo {author}
  {\bibfnamefont {L.}~\bibnamefont {Lu}},\ and\ \bibinfo {author}
  {\bibfnamefont {X.}~\bibnamefont {Zhang}},\ }\bibfield  {title} {\bibinfo
  {title} {Infrared topological plasmons in graphene},\ }\href
  {https://doi.org/10.1103/PhysRevLett.118.245301} {\bibfield  {journal}
  {\bibinfo  {journal} {Phys. Rev. Lett.}\ }\textbf {\bibinfo {volume} {118}},\
  \bibinfo {pages} {245301} (\bibinfo {year} {2017})}\BibitemShut {NoStop}%
\bibitem [{\citenamefont {Pan}\ \emph {et~al.}(2017)\citenamefont {Pan},
  \citenamefont {Yu}, \citenamefont {Xu},\ and\ \citenamefont {Garc{\'i}a~de
  Abajo}}]{pan2017topologically}%
  \BibitemOpen
  \bibfield  {author} {\bibinfo {author} {\bibfnamefont {D.}~\bibnamefont
  {Pan}}, \bibinfo {author} {\bibfnamefont {R.}~\bibnamefont {Yu}}, \bibinfo
  {author} {\bibfnamefont {H.}~\bibnamefont {Xu}},\ and\ \bibinfo {author}
  {\bibfnamefont {F.~J.}\ \bibnamefont {Garc{\'i}a~de Abajo}},\ }\bibfield
  {title} {\bibinfo {title} {Topologically protected dirac plasmons in a
  graphene superlattice},\ }\href {https://doi.org/10.1038/s41467-017-01205-z}
  {\bibfield  {journal} {\bibinfo  {journal} {Nature Communications}\ }\textbf
  {\bibinfo {volume} {8}},\ \bibinfo {pages} {1243} (\bibinfo {year}
  {2017})}\BibitemShut {NoStop}%
\bibitem [{\citenamefont {Makwana}\ and\ \citenamefont
  {Craster}(2018)}]{makwana2018geometrically}%
  \BibitemOpen
  \bibfield  {author} {\bibinfo {author} {\bibfnamefont {M.~P.}\ \bibnamefont
  {Makwana}}\ and\ \bibinfo {author} {\bibfnamefont {R.~V.}\ \bibnamefont
  {Craster}},\ }\bibfield  {title} {\bibinfo {title} {Geometrically navigating
  topological plate modes around gentle and sharp bends},\ }\href
  {https://doi.org/10.1103/PhysRevB.98.184105} {\bibfield  {journal} {\bibinfo
  {journal} {Phys. Rev. B}\ }\textbf {\bibinfo {volume} {98}},\ \bibinfo
  {pages} {184105} (\bibinfo {year} {2018})}\BibitemShut {NoStop}%
\bibitem [{\citenamefont {Wong}\ \emph {et~al.}(2020)\citenamefont {Wong},
  \citenamefont {Saba}, \citenamefont {Hess},\ and\ \citenamefont
  {Oh}}]{wong2020gapless}%
  \BibitemOpen
  \bibfield  {author} {\bibinfo {author} {\bibfnamefont {S.}~\bibnamefont
  {Wong}}, \bibinfo {author} {\bibfnamefont {M.}~\bibnamefont {Saba}}, \bibinfo
  {author} {\bibfnamefont {O.}~\bibnamefont {Hess}},\ and\ \bibinfo {author}
  {\bibfnamefont {S.~S.}\ \bibnamefont {Oh}},\ }\bibfield  {title} {\bibinfo
  {title} {Gapless unidirectional photonic transport using all-dielectric
  kagome lattices},\ }\href {https://doi.org/10.1103/PhysRevResearch.2.012011}
  {\bibfield  {journal} {\bibinfo  {journal} {Phys. Rev. Research}\ }\textbf
  {\bibinfo {volume} {2}},\ \bibinfo {pages} {012011} (\bibinfo {year}
  {2020})}\BibitemShut {NoStop}%
\bibitem [{\citenamefont {Proctor}\ \emph
  {et~al.}(2020{\natexlab{a}})\citenamefont {Proctor}, \citenamefont
  {Huidobro}, \citenamefont {Maier}, \citenamefont {Craster},\ and\
  \citenamefont {Makwana}}]{proctor2020manipulating}%
  \BibitemOpen
  \bibfield  {author} {\bibinfo {author} {\bibfnamefont {M.}~\bibnamefont
  {Proctor}}, \bibinfo {author} {\bibfnamefont {P.~A.}\ \bibnamefont
  {Huidobro}}, \bibinfo {author} {\bibfnamefont {S.~A.}\ \bibnamefont {Maier}},
  \bibinfo {author} {\bibfnamefont {R.~V.}\ \bibnamefont {Craster}},\ and\
  \bibinfo {author} {\bibfnamefont {M.~P.}\ \bibnamefont {Makwana}},\
  }\bibfield  {title} {\bibinfo {title} {Manipulating topological valley modes
  in plasmonic metasurfaces},\ }\href
  {https://doi.org/10.1515/nanoph-2019-0408} {\bibfield  {journal} {\bibinfo
  {journal} {Nanophotonics}\ }\textbf {\bibinfo {volume} {9}},\ \bibinfo
  {pages} {657 } (\bibinfo {year} {2020}{\natexlab{a}})}\BibitemShut {NoStop}%
\bibitem [{\citenamefont {Saba}\ \emph {et~al.}(2020)\citenamefont {Saba},
  \citenamefont {Wong}, \citenamefont {Elman}, \citenamefont {Oh},\ and\
  \citenamefont {Hess}}]{saba2020nature}%
  \BibitemOpen
  \bibfield  {author} {\bibinfo {author} {\bibfnamefont {M.}~\bibnamefont
  {Saba}}, \bibinfo {author} {\bibfnamefont {S.}~\bibnamefont {Wong}}, \bibinfo
  {author} {\bibfnamefont {M.}~\bibnamefont {Elman}}, \bibinfo {author}
  {\bibfnamefont {S.~S.}\ \bibnamefont {Oh}},\ and\ \bibinfo {author}
  {\bibfnamefont {O.}~\bibnamefont {Hess}},\ }\bibfield  {title} {\bibinfo
  {title} {Nature of topological protection in photonic spin and valley hall
  insulators},\ }\href {https://doi.org/10.1103/PhysRevB.101.054307} {\bibfield
   {journal} {\bibinfo  {journal} {Phys. Rev. B}\ }\textbf {\bibinfo {volume}
  {101}},\ \bibinfo {pages} {054307} (\bibinfo {year} {2020})}\BibitemShut
  {NoStop}%
\bibitem [{\citenamefont {Orazbayev}\ and\ \citenamefont
  {Fleury}(2019)}]{orazbayev2019quantitative}%
  \BibitemOpen
  \bibfield  {author} {\bibinfo {author} {\bibfnamefont {B.}~\bibnamefont
  {Orazbayev}}\ and\ \bibinfo {author} {\bibfnamefont {R.}~\bibnamefont
  {Fleury}},\ }\bibfield  {title} {\bibinfo {title} {Quantitative robustness
  analysis of topological edge modes in c6 and valley-hall metamaterial
  waveguides},\ }\href {https://doi.org/10.1515/nanoph-2019-0137} {\bibfield
  {journal} {\bibinfo  {journal} {Nanophotonics}\ }\textbf {\bibinfo {volume}
  {8}},\ \bibinfo {pages} {1433 } (\bibinfo {year} {2019})}\BibitemShut
  {NoStop}%
\bibitem [{\citenamefont {Wu}\ and\ \citenamefont {Hu}(2015)}]{wu2015scheme}%
  \BibitemOpen
  \bibfield  {author} {\bibinfo {author} {\bibfnamefont {L.-H.}\ \bibnamefont
  {Wu}}\ and\ \bibinfo {author} {\bibfnamefont {X.}~\bibnamefont {Hu}},\
  }\bibfield  {title} {\bibinfo {title} {Scheme for achieving a topological
  photonic crystal by using dielectric material},\ }\href
  {https://doi.org/10.1103/PhysRevLett.114.223901} {\bibfield  {journal}
  {\bibinfo  {journal} {Phys. Rev. Lett.}\ }\textbf {\bibinfo {volume} {114}},\
  \bibinfo {pages} {223901} (\bibinfo {year} {2015})}\BibitemShut {NoStop}%
\bibitem [{\citenamefont {de~Paz}\ \emph {et~al.}(2019)\citenamefont {de~Paz},
  \citenamefont {Vergniory}, \citenamefont {Bercioux}, \citenamefont
  {Garc\'{\i}a-Etxarri},\ and\ \citenamefont {Bradlyn}}]{depaz2019engineering}%
  \BibitemOpen
  \bibfield  {author} {\bibinfo {author} {\bibfnamefont {M.~B.}\ \bibnamefont
  {de~Paz}}, \bibinfo {author} {\bibfnamefont {M.~G.}\ \bibnamefont
  {Vergniory}}, \bibinfo {author} {\bibfnamefont {D.}~\bibnamefont {Bercioux}},
  \bibinfo {author} {\bibfnamefont {A.}~\bibnamefont {Garc\'{\i}a-Etxarri}},\
  and\ \bibinfo {author} {\bibfnamefont {B.}~\bibnamefont {Bradlyn}},\
  }\bibfield  {title} {\bibinfo {title} {Engineering fragile topology in
  photonic crystals: Topological quantum chemistry of light},\ }\href
  {https://doi.org/10.1103/PhysRevResearch.1.032005} {\bibfield  {journal}
  {\bibinfo  {journal} {Phys. Rev. Research}\ }\textbf {\bibinfo {volume}
  {1}},\ \bibinfo {pages} {032005} (\bibinfo {year} {2019})}\BibitemShut
  {NoStop}%
\bibitem [{\citenamefont {Proctor}\ \emph
  {et~al.}(2020{\natexlab{b}})\citenamefont {Proctor}, \citenamefont
  {Huidobro}, \citenamefont {Bradlyn}, \citenamefont {de~Paz}, \citenamefont
  {Vergniory}, \citenamefont {Bercioux},\ and\ \citenamefont
  {Garcia-Etxarri}}]{proctor2020robustness}%
  \BibitemOpen
  \bibfield  {author} {\bibinfo {author} {\bibfnamefont {M.}~\bibnamefont
  {Proctor}}, \bibinfo {author} {\bibfnamefont {P.~A.}\ \bibnamefont
  {Huidobro}}, \bibinfo {author} {\bibfnamefont {B.}~\bibnamefont {Bradlyn}},
  \bibinfo {author} {\bibfnamefont {M.~B.}\ \bibnamefont {de~Paz}}, \bibinfo
  {author} {\bibfnamefont {M.~G.}\ \bibnamefont {Vergniory}}, \bibinfo {author}
  {\bibfnamefont {D.}~\bibnamefont {Bercioux}},\ and\ \bibinfo {author}
  {\bibfnamefont {A.}~\bibnamefont {Garcia-Etxarri}},\ }\href@noop {} {\bibinfo
  {title} {On the robustness of topological corner modes in photonic crystals}}
  (\bibinfo {year} {2020}{\natexlab{b}}),\ \Eprint
  {https://arxiv.org/abs/2007.10624} {arXiv:2007.10624 [cond-mat.mes-hall]}
  \BibitemShut {NoStop}%
\bibitem [{\citenamefont {Proctor}\ \emph {et~al.}(2019)\citenamefont
  {Proctor}, \citenamefont {Craster}, \citenamefont {Maier}, \citenamefont
  {Giannini},\ and\ \citenamefont {Huidobro}}]{proctor2019exciting}%
  \BibitemOpen
  \bibfield  {author} {\bibinfo {author} {\bibfnamefont {M.}~\bibnamefont
  {Proctor}}, \bibinfo {author} {\bibfnamefont {R.~V.}\ \bibnamefont
  {Craster}}, \bibinfo {author} {\bibfnamefont {S.~A.}\ \bibnamefont {Maier}},
  \bibinfo {author} {\bibfnamefont {V.}~\bibnamefont {Giannini}},\ and\
  \bibinfo {author} {\bibfnamefont {P.~A.}\ \bibnamefont {Huidobro}},\
  }\bibfield  {title} {\bibinfo {title} {Exciting pseudospin-dependent edge
  states in plasmonic metasurfaces},\ }\href
  {https://doi.org/10.1021/acsphotonics.9b01192} {\bibfield  {journal}
  {\bibinfo  {journal} {ACS Photonics}\ }\textbf {\bibinfo {volume} {6}},\
  \bibinfo {pages} {2985} (\bibinfo {year} {2019})}\BibitemShut {NoStop}%
\bibitem [{\citenamefont {Oh}\ \emph {et~al.}(2018)\citenamefont {Oh},
  \citenamefont {Lang}, \citenamefont {Beggs}, \citenamefont {Huffaker},
  \citenamefont {Saba},\ and\ \citenamefont {Hess}}]{oh2018chiral}%
  \BibitemOpen
  \bibfield  {author} {\bibinfo {author} {\bibfnamefont {S.~S.}\ \bibnamefont
  {Oh}}, \bibinfo {author} {\bibfnamefont {B.}~\bibnamefont {Lang}}, \bibinfo
  {author} {\bibfnamefont {D.~M.}\ \bibnamefont {Beggs}}, \bibinfo {author}
  {\bibfnamefont {D.~L.}\ \bibnamefont {Huffaker}}, \bibinfo {author}
  {\bibfnamefont {M.}~\bibnamefont {Saba}},\ and\ \bibinfo {author}
  {\bibfnamefont {O.}~\bibnamefont {Hess}},\ }\bibfield  {title} {\bibinfo
  {title} {Chiral light-matter interaction in dielectric photonic topological
  insulators},\ }in\ \href {https://doi.org/10.1364/CLEOPR.2018.Th4H.5} {\emph
  {\bibinfo {booktitle} {CLEO Pacific Rim Conference 2018}}}\ (\bibinfo
  {publisher} {Optical Society of America},\ \bibinfo {year} {2018})\ p.\
  \bibinfo {pages} {Th4H.5}\BibitemShut {NoStop}%
\bibitem [{\citenamefont {Smirnova}\ \emph {et~al.}(2019)\citenamefont
  {Smirnova}, \citenamefont {Kruk}, \citenamefont {Leykam}, \citenamefont
  {Melik-Gaykazyan}, \citenamefont {Choi},\ and\ \citenamefont
  {Kivshar}}]{smirnova2019third}%
  \BibitemOpen
  \bibfield  {author} {\bibinfo {author} {\bibfnamefont {D.}~\bibnamefont
  {Smirnova}}, \bibinfo {author} {\bibfnamefont {S.}~\bibnamefont {Kruk}},
  \bibinfo {author} {\bibfnamefont {D.}~\bibnamefont {Leykam}}, \bibinfo
  {author} {\bibfnamefont {E.}~\bibnamefont {Melik-Gaykazyan}}, \bibinfo
  {author} {\bibfnamefont {D.-Y.}\ \bibnamefont {Choi}},\ and\ \bibinfo
  {author} {\bibfnamefont {Y.}~\bibnamefont {Kivshar}},\ }\bibfield  {title}
  {\bibinfo {title} {Third-harmonic generation in photonic topological
  metasurfaces},\ }\href {https://doi.org/10.1103/PhysRevLett.123.103901}
  {\bibfield  {journal} {\bibinfo  {journal} {Phys. Rev. Lett.}\ }\textbf
  {\bibinfo {volume} {123}},\ \bibinfo {pages} {103901} (\bibinfo {year}
  {2019})}\BibitemShut {NoStop}%
\bibitem [{\citenamefont {Barik}\ \emph {et~al.}(2016)\citenamefont {Barik},
  \citenamefont {Miyake}, \citenamefont {DeGottardi}, \citenamefont {Waks},\
  and\ \citenamefont {Hafezi}}]{barik2016two}%
  \BibitemOpen
  \bibfield  {author} {\bibinfo {author} {\bibfnamefont {S.}~\bibnamefont
  {Barik}}, \bibinfo {author} {\bibfnamefont {H.}~\bibnamefont {Miyake}},
  \bibinfo {author} {\bibfnamefont {W.}~\bibnamefont {DeGottardi}}, \bibinfo
  {author} {\bibfnamefont {E.}~\bibnamefont {Waks}},\ and\ \bibinfo {author}
  {\bibfnamefont {M.}~\bibnamefont {Hafezi}},\ }\bibfield  {title} {\bibinfo
  {title} {Two-dimensionally confined topological edge states in photonic
  crystals},\ }\href {https://doi.org/10.1088/1367-2630/18/11/113013}
  {\bibfield  {journal} {\bibinfo  {journal} {New Journal of Physics}\ }\textbf
  {\bibinfo {volume} {18}},\ \bibinfo {pages} {113013} (\bibinfo {year}
  {2016})}\BibitemShut {NoStop}%
\bibitem [{\citenamefont {Barik}\ \emph {et~al.}(2018)\citenamefont {Barik},
  \citenamefont {Karasahin}, \citenamefont {Flower}, \citenamefont {Cai},
  \citenamefont {Miyake}, \citenamefont {DeGottardi}, \citenamefont {Hafezi},\
  and\ \citenamefont {Waks}}]{barik2018topological}%
  \BibitemOpen
  \bibfield  {author} {\bibinfo {author} {\bibfnamefont {S.}~\bibnamefont
  {Barik}}, \bibinfo {author} {\bibfnamefont {A.}~\bibnamefont {Karasahin}},
  \bibinfo {author} {\bibfnamefont {C.}~\bibnamefont {Flower}}, \bibinfo
  {author} {\bibfnamefont {T.}~\bibnamefont {Cai}}, \bibinfo {author}
  {\bibfnamefont {H.}~\bibnamefont {Miyake}}, \bibinfo {author} {\bibfnamefont
  {W.}~\bibnamefont {DeGottardi}}, \bibinfo {author} {\bibfnamefont
  {M.}~\bibnamefont {Hafezi}},\ and\ \bibinfo {author} {\bibfnamefont
  {E.}~\bibnamefont {Waks}},\ }\bibfield  {title} {\bibinfo {title} {A
  topological quantum optics interface},\ }\href
  {https://doi.org/10.1126/science.aaq0327} {\bibfield  {journal} {\bibinfo
  {journal} {Science}\ }\textbf {\bibinfo {volume} {359}},\ \bibinfo {pages}
  {666} (\bibinfo {year} {2018})}\BibitemShut {NoStop}%
\bibitem [{\citenamefont {Yves}\ \emph {et~al.}(2017)\citenamefont {Yves},
  \citenamefont {Fleury}, \citenamefont {Berthelot}, \citenamefont {Fink},
  \citenamefont {Lemoult},\ and\ \citenamefont
  {Lerosey}}]{yves2017crystalline}%
  \BibitemOpen
  \bibfield  {author} {\bibinfo {author} {\bibfnamefont {S.}~\bibnamefont
  {Yves}}, \bibinfo {author} {\bibfnamefont {R.}~\bibnamefont {Fleury}},
  \bibinfo {author} {\bibfnamefont {T.}~\bibnamefont {Berthelot}}, \bibinfo
  {author} {\bibfnamefont {M.}~\bibnamefont {Fink}}, \bibinfo {author}
  {\bibfnamefont {F.}~\bibnamefont {Lemoult}},\ and\ \bibinfo {author}
  {\bibfnamefont {G.}~\bibnamefont {Lerosey}},\ }\bibfield  {title} {\bibinfo
  {title} {Crystalline metamaterials for topological properties at
  subwavelength scales},\ }\href {https://doi.org/10.1038/ncomms16023}
  {\bibfield  {journal} {\bibinfo  {journal} {Nature Communications}\ }\textbf
  {\bibinfo {volume} {8}},\ \bibinfo {pages} {16023} (\bibinfo {year}
  {2017})}\BibitemShut {NoStop}%
\bibitem [{\citenamefont {Parappurath}\ \emph {et~al.}(2020)\citenamefont
  {Parappurath}, \citenamefont {Alpeggiani}, \citenamefont {Kuipers},\ and\
  \citenamefont {Verhagen}}]{parappurath2020topological}%
  \BibitemOpen
  \bibfield  {author} {\bibinfo {author} {\bibfnamefont {N.}~\bibnamefont
  {Parappurath}}, \bibinfo {author} {\bibfnamefont {F.}~\bibnamefont
  {Alpeggiani}}, \bibinfo {author} {\bibfnamefont {L.}~\bibnamefont
  {Kuipers}},\ and\ \bibinfo {author} {\bibfnamefont {E.}~\bibnamefont
  {Verhagen}},\ }\bibfield  {title} {\bibinfo {title} {Direct observation of
  topological edge states in silicon photonic crystals: Spin, dispersion, and
  chiral routing},\ }\bibfield  {journal} {\bibinfo  {journal} {Science
  Advances}\ }\textbf {\bibinfo {volume} {6}},\ \href
  {https://doi.org/10.1126/sciadv.aaw4137} {10.1126/sciadv.aaw4137} (\bibinfo
  {year} {2020})\BibitemShut {NoStop}%
\bibitem [{\citenamefont {Liu}\ \emph {et~al.}(2020)\citenamefont {Liu},
  \citenamefont {Hwang}, \citenamefont {Ji}, \citenamefont {Wang},
  \citenamefont {Modi},\ and\ \citenamefont {Agarwal}}]{Liu2020photonic}%
  \BibitemOpen
  \bibfield  {author} {\bibinfo {author} {\bibfnamefont {W.}~\bibnamefont
  {Liu}}, \bibinfo {author} {\bibfnamefont {M.}~\bibnamefont {Hwang}}, \bibinfo
  {author} {\bibfnamefont {Z.}~\bibnamefont {Ji}}, \bibinfo {author}
  {\bibfnamefont {Y.}~\bibnamefont {Wang}}, \bibinfo {author} {\bibfnamefont
  {G.}~\bibnamefont {Modi}},\ and\ \bibinfo {author} {\bibfnamefont
  {R.}~\bibnamefont {Agarwal}},\ }\bibfield  {title} {\bibinfo {title} {$z_2$
  photonic topological insulators in the visible wavelength range for robust
  nanoscale photonics},\ }\href {https://doi.org/10.1021/acs.nanolett.9b04813}
  {\bibfield  {journal} {\bibinfo  {journal} {Nano Letters}\ }\textbf {\bibinfo
  {volume} {20}},\ \bibinfo {pages} {1329–1335} (\bibinfo {year}
  {2020})}\BibitemShut {NoStop}%
\bibitem [{\citenamefont {Maier}(2007)}]{maier2007plasmonics}%
  \BibitemOpen
  \bibfield  {author} {\bibinfo {author} {\bibfnamefont {S.~A.}\ \bibnamefont
  {Maier}},\ }\href@noop {} {\emph {\bibinfo {title} {Plasmonics:
  {F}undamentals and {A}pplications}}}\ (\bibinfo  {publisher} {Springer
  Science \& Business Media},\ \bibinfo {year} {2007})\BibitemShut {NoStop}%
\bibitem [{\citenamefont {Weber}\ and\ \citenamefont
  {Ford}(2004)}]{weber2004propagation}%
  \BibitemOpen
  \bibfield  {author} {\bibinfo {author} {\bibfnamefont {W.~H.}\ \bibnamefont
  {Weber}}\ and\ \bibinfo {author} {\bibfnamefont {G.~W.}\ \bibnamefont
  {Ford}},\ }\bibfield  {title} {\bibinfo {title} {Propagation of optical
  excitations by dipolar interactions in metal nanoparticle chains},\ }\href
  {https://doi.org/10.1103/PhysRevB.70.125429} {\bibfield  {journal} {\bibinfo
  {journal} {Phys. Rev. B}\ }\textbf {\bibinfo {volume} {70}},\ \bibinfo
  {pages} {125429} (\bibinfo {year} {2004})}\BibitemShut {NoStop}%
\bibitem [{\citenamefont {Moroz}(2009)}]{moroz2009depolarization}%
  \BibitemOpen
  \bibfield  {author} {\bibinfo {author} {\bibfnamefont {A.}~\bibnamefont
  {Moroz}},\ }\bibfield  {title} {\bibinfo {title} {Depolarization field of
  spheroidal particles},\ }\href {https://doi.org/10.1364/JOSAB.26.000517}
  {\bibfield  {journal} {\bibinfo  {journal} {J. Opt. Soc. Am. B}\ }\textbf
  {\bibinfo {volume} {26}},\ \bibinfo {pages} {517} (\bibinfo {year}
  {2009})}\BibitemShut {NoStop}%
\bibitem [{\citenamefont {Meier}\ and\ \citenamefont
  {Wokaun}(1983)}]{meier1983depolarization}%
  \BibitemOpen
  \bibfield  {author} {\bibinfo {author} {\bibfnamefont {M.}~\bibnamefont
  {Meier}}\ and\ \bibinfo {author} {\bibfnamefont {A.}~\bibnamefont {Wokaun}},\
  }\bibfield  {title} {\bibinfo {title} {Enhanced fields on large metal
  particles: dynamic depolarization},\ }\href
  {https://doi.org/10.1364/OL.8.000581} {\bibfield  {journal} {\bibinfo
  {journal} {Opt. Lett.}\ }\textbf {\bibinfo {volume} {8}},\ \bibinfo {pages}
  {581} (\bibinfo {year} {1983})}\BibitemShut {NoStop}%
\bibitem [{\citenamefont {Yang}\ \emph {et~al.}(2015)\citenamefont {Yang},
  \citenamefont {D'Archangel}, \citenamefont {Sundheimer}, \citenamefont
  {Tucker}, \citenamefont {Boreman},\ and\ \citenamefont
  {Raschke}}]{yang2015optical}%
  \BibitemOpen
  \bibfield  {author} {\bibinfo {author} {\bibfnamefont {H.~U.}\ \bibnamefont
  {Yang}}, \bibinfo {author} {\bibfnamefont {J.}~\bibnamefont {D'Archangel}},
  \bibinfo {author} {\bibfnamefont {M.~L.}\ \bibnamefont {Sundheimer}},
  \bibinfo {author} {\bibfnamefont {E.}~\bibnamefont {Tucker}}, \bibinfo
  {author} {\bibfnamefont {G.~D.}\ \bibnamefont {Boreman}},\ and\ \bibinfo
  {author} {\bibfnamefont {M.~B.}\ \bibnamefont {Raschke}},\ }\bibfield
  {title} {\bibinfo {title} {Optical dielectric function of silver},\ }\href
  {https://doi.org/10.1103/PhysRevB.91.235137} {\bibfield  {journal} {\bibinfo
  {journal} {Phys. Rev. B}\ }\textbf {\bibinfo {volume} {91}},\ \bibinfo
  {pages} {235137} (\bibinfo {year} {2015})}\BibitemShut {NoStop}%
\bibitem [{\citenamefont {Garc\'{\i}a~de Abajo}(2007)}]{abajo2007colloquium}%
  \BibitemOpen
  \bibfield  {author} {\bibinfo {author} {\bibfnamefont {F.~J.}\ \bibnamefont
  {Garc\'{\i}a~de Abajo}},\ }\bibfield  {title} {\bibinfo {title} {Colloquium:
  Light scattering by particle and hole arrays},\ }\href
  {https://doi.org/10.1103/RevModPhys.79.1267} {\bibfield  {journal} {\bibinfo
  {journal} {Rev. Mod. Phys.}\ }\textbf {\bibinfo {volume} {79}},\ \bibinfo
  {pages} {1267} (\bibinfo {year} {2007})}\BibitemShut {NoStop}%
\bibitem [{\citenamefont {Wang}\ \emph {et~al.}(2016)\citenamefont {Wang},
  \citenamefont {Zhang}, \citenamefont {Xiao}, \citenamefont {Han},
  \citenamefont {Chan},\ and\ \citenamefont {Wen}}]{wang2016existence}%
  \BibitemOpen
  \bibfield  {author} {\bibinfo {author} {\bibfnamefont {L.}~\bibnamefont
  {Wang}}, \bibinfo {author} {\bibfnamefont {R.-Y.}\ \bibnamefont {Zhang}},
  \bibinfo {author} {\bibfnamefont {M.}~\bibnamefont {Xiao}}, \bibinfo {author}
  {\bibfnamefont {D.}~\bibnamefont {Han}}, \bibinfo {author} {\bibfnamefont
  {C.~T.}\ \bibnamefont {Chan}},\ and\ \bibinfo {author} {\bibfnamefont
  {W.}~\bibnamefont {Wen}},\ }\bibfield  {title} {\bibinfo {title} {The
  existence of topological edge states in honeycomb plasmonic lattices},\
  }\href {https://doi.org/10.1088/1367-2630/18/10/103029} {\bibfield  {journal}
  {\bibinfo  {journal} {New Journal of Physics}\ }\textbf {\bibinfo {volume}
  {18}},\ \bibinfo {pages} {103029} (\bibinfo {year} {2016})}\BibitemShut
  {NoStop}%
\bibitem [{\citenamefont {Linton}(2010)}]{linton2010lattice}%
  \BibitemOpen
  \bibfield  {author} {\bibinfo {author} {\bibfnamefont {C.~M.}\ \bibnamefont
  {Linton}},\ }\bibfield  {title} {\bibinfo {title} {Lattice sums for the
  helmholtz equation},\ }\href {https://doi.org/10.1137/09075130X} {\bibfield
  {journal} {\bibinfo  {journal} {SIAM Review}\ }\textbf {\bibinfo {volume}
  {52}},\ \bibinfo {pages} {630} (\bibinfo {year} {2010})}\BibitemShut
  {NoStop}%
\bibitem [{\citenamefont {{Kolkowski}}\ and\ \citenamefont
  {{Koenderink}}(2020)}]{Kolkowski2020lattice}%
  \BibitemOpen
  \bibfield  {author} {\bibinfo {author} {\bibfnamefont {R.}~\bibnamefont
  {{Kolkowski}}}\ and\ \bibinfo {author} {\bibfnamefont {A.~F.}\ \bibnamefont
  {{Koenderink}}},\ }\bibfield  {title} {\bibinfo {title} {Lattice resonances
  in optical metasurfaces with gain and loss},\ }\href
  {https://doi.org/10.1109/JPROC.2019.2939396} {\bibfield  {journal} {\bibinfo
  {journal} {Proceedings of the IEEE}\ }\textbf {\bibinfo {volume} {108}},\
  \bibinfo {pages} {795} (\bibinfo {year} {2020})}\BibitemShut {NoStop}%
\bibitem [{\citenamefont {Koenderink}\ and\ \citenamefont
  {Polman}(2006)}]{koenderink2006complex}%
  \BibitemOpen
  \bibfield  {author} {\bibinfo {author} {\bibfnamefont {A.~F.}\ \bibnamefont
  {Koenderink}}\ and\ \bibinfo {author} {\bibfnamefont {A.}~\bibnamefont
  {Polman}},\ }\bibfield  {title} {\bibinfo {title} {Complex response and
  polariton-like dispersion splitting in periodic metal nanoparticle chains},\
  }\href {https://doi.org/10.1103/PhysRevB.74.033402} {\bibfield  {journal}
  {\bibinfo  {journal} {Phys. Rev. B}\ }\textbf {\bibinfo {volume} {74}},\
  \bibinfo {pages} {033402} (\bibinfo {year} {2006})}\BibitemShut {NoStop}%
\bibitem [{\citenamefont {Zhen}\ \emph {et~al.}(2008)\citenamefont {Zhen},
  \citenamefont {Fung},\ and\ \citenamefont {Chan}}]{zhen2008collective}%
  \BibitemOpen
  \bibfield  {author} {\bibinfo {author} {\bibfnamefont {Y.-R.}\ \bibnamefont
  {Zhen}}, \bibinfo {author} {\bibfnamefont {K.~H.}\ \bibnamefont {Fung}},\
  and\ \bibinfo {author} {\bibfnamefont {C.~T.}\ \bibnamefont {Chan}},\
  }\bibfield  {title} {\bibinfo {title} {Collective plasmonic modes in
  two-dimensional periodic arrays of metal nanoparticles},\ }\href
  {https://doi.org/10.1103/PhysRevB.78.035419} {\bibfield  {journal} {\bibinfo
  {journal} {Phys. Rev. B}\ }\textbf {\bibinfo {volume} {78}},\ \bibinfo
  {pages} {035419} (\bibinfo {year} {2008})}\BibitemShut {NoStop}%
\bibitem [{\citenamefont {Pocock}\ \emph {et~al.}(2018)\citenamefont {Pocock},
  \citenamefont {Xiao}, \citenamefont {Huidobro},\ and\ \citenamefont
  {Giannini}}]{pocock2018topological}%
  \BibitemOpen
  \bibfield  {author} {\bibinfo {author} {\bibfnamefont {S.~R.}\ \bibnamefont
  {Pocock}}, \bibinfo {author} {\bibfnamefont {X.}~\bibnamefont {Xiao}},
  \bibinfo {author} {\bibfnamefont {P.~A.}\ \bibnamefont {Huidobro}},\ and\
  \bibinfo {author} {\bibfnamefont {V.}~\bibnamefont {Giannini}},\ }\bibfield
  {title} {\bibinfo {title} {Topological plasmonic chain with retardation and
  radiative effects},\ }\href {https://doi.org/10.1021/acsphotonics.8b00117}
  {\bibfield  {journal} {\bibinfo  {journal} {ACS Photonics}\ }\textbf
  {\bibinfo {volume} {5}},\ \bibinfo {pages} {2271} (\bibinfo {year}
  {2018})}\BibitemShut {NoStop}%
\bibitem [{\citenamefont {Pocock}\ \emph {et~al.}(2019)\citenamefont {Pocock},
  \citenamefont {Huidobro},\ and\ \citenamefont {Giannini}}]{pocock2019bulk}%
  \BibitemOpen
  \bibfield  {author} {\bibinfo {author} {\bibfnamefont {S.~R.}\ \bibnamefont
  {Pocock}}, \bibinfo {author} {\bibfnamefont {P.~A.}\ \bibnamefont
  {Huidobro}},\ and\ \bibinfo {author} {\bibfnamefont {V.}~\bibnamefont
  {Giannini}},\ }\bibfield  {title} {\bibinfo {title} {Bulk-edge correspondence
  and long-range hopping in the topological plasmonic chain},\ }\href
  {https://doi.org/10.1515/nanoph-2019-0033} {\bibfield  {journal} {\bibinfo
  {journal} {Nanophotonics}\ }\textbf {\bibinfo {volume} {8}},\ \bibinfo
  {pages} {1337 } (\bibinfo {year} {2019})}\BibitemShut {NoStop}%
\bibitem [{\citenamefont {Merchiers}\ \emph {et~al.}(2007)\citenamefont
  {Merchiers}, \citenamefont {Moreno}, \citenamefont {Gonz\'alez},\ and\
  \citenamefont {Saiz}}]{Merchiers2007Light}%
  \BibitemOpen
  \bibfield  {author} {\bibinfo {author} {\bibfnamefont {O.}~\bibnamefont
  {Merchiers}}, \bibinfo {author} {\bibfnamefont {F.}~\bibnamefont {Moreno}},
  \bibinfo {author} {\bibfnamefont {F.}~\bibnamefont {Gonz\'alez}},\ and\
  \bibinfo {author} {\bibfnamefont {J.~M.}\ \bibnamefont {Saiz}},\ }\bibfield
  {title} {\bibinfo {title} {Light scattering by an ensemble of interacting
  dipolar particles with both electric and magnetic polarizabilities},\ }\href
  {https://doi.org/10.1103/PhysRevA.76.043834} {\bibfield  {journal} {\bibinfo
  {journal} {Phys. Rev. A}\ }\textbf {\bibinfo {volume} {76}},\ \bibinfo
  {pages} {043834} (\bibinfo {year} {2007})}\BibitemShut {NoStop}%
\bibitem [{\citenamefont {Gorlach}\ \emph {et~al.}(2018)\citenamefont
  {Gorlach}, \citenamefont {Ni}, \citenamefont {Smirnova}, \citenamefont
  {Korobkin}, \citenamefont {Zhirihin}, \citenamefont {Slobozhanyuk},
  \citenamefont {Belov}, \citenamefont {Al{\`u}},\ and\ \citenamefont
  {Khanikaev}}]{gorlach2018far}%
  \BibitemOpen
  \bibfield  {author} {\bibinfo {author} {\bibfnamefont {M.~A.}\ \bibnamefont
  {Gorlach}}, \bibinfo {author} {\bibfnamefont {X.}~\bibnamefont {Ni}},
  \bibinfo {author} {\bibfnamefont {D.~A.}\ \bibnamefont {Smirnova}}, \bibinfo
  {author} {\bibfnamefont {D.}~\bibnamefont {Korobkin}}, \bibinfo {author}
  {\bibfnamefont {D.}~\bibnamefont {Zhirihin}}, \bibinfo {author}
  {\bibfnamefont {A.~P.}\ \bibnamefont {Slobozhanyuk}}, \bibinfo {author}
  {\bibfnamefont {P.~A.}\ \bibnamefont {Belov}}, \bibinfo {author}
  {\bibfnamefont {A.}~\bibnamefont {Al{\`u}}},\ and\ \bibinfo {author}
  {\bibfnamefont {A.~B.}\ \bibnamefont {Khanikaev}},\ }\bibfield  {title}
  {\bibinfo {title} {Far-field probing of leaky topological states in
  all-dielectric metasurfaces},\ }\href
  {https://doi.org/10.1038/s41467-018-03330-9} {\bibfield  {journal} {\bibinfo
  {journal} {Nature Communications}\ }\textbf {\bibinfo {volume} {9}},\
  \bibinfo {pages} {909} (\bibinfo {year} {2018})}\BibitemShut {NoStop}%
\bibitem [{\citenamefont {Blanco~de Paz}\ \emph {et~al.}(2020)\citenamefont
  {Blanco~de Paz}, \citenamefont {Devescovi}, \citenamefont {Giedke},
  \citenamefont {Saenz}, \citenamefont {Vergniory}, \citenamefont {Bradlyn},
  \citenamefont {Bercioux},\ and\ \citenamefont
  {García-Etxarri}}]{depaz2020tutorial}%
  \BibitemOpen
  \bibfield  {author} {\bibinfo {author} {\bibfnamefont {M.}~\bibnamefont
  {Blanco~de Paz}}, \bibinfo {author} {\bibfnamefont {C.}~\bibnamefont
  {Devescovi}}, \bibinfo {author} {\bibfnamefont {G.}~\bibnamefont {Giedke}},
  \bibinfo {author} {\bibfnamefont {J.~J.}\ \bibnamefont {Saenz}}, \bibinfo
  {author} {\bibfnamefont {M.~G.}\ \bibnamefont {Vergniory}}, \bibinfo {author}
  {\bibfnamefont {B.}~\bibnamefont {Bradlyn}}, \bibinfo {author} {\bibfnamefont
  {D.}~\bibnamefont {Bercioux}},\ and\ \bibinfo {author} {\bibfnamefont
  {A.}~\bibnamefont {García-Etxarri}},\ }\bibfield  {title} {\bibinfo {title}
  {Tutorial: Computing topological invariants in 2d photonic crystals},\ }\href
  {https://doi.org/10.1002/qute.201900117} {\bibfield  {journal} {\bibinfo
  {journal} {Advanced Quantum Technologies}\ }\textbf {\bibinfo {volume} {3}},\
  \bibinfo {pages} {1900117} (\bibinfo {year} {2020})}\BibitemShut {NoStop}%
\bibitem [{\citenamefont {Lodahl}\ \emph {et~al.}(2017)\citenamefont {Lodahl},
  \citenamefont {Mahmoodian}, \citenamefont {Stobbe}, \citenamefont
  {Rauschenbeutel}, \citenamefont {Schneeweiss}, \citenamefont {Volz},
  \citenamefont {Pichler},\ and\ \citenamefont {Zoller}}]{lodahl2017chiral}%
  \BibitemOpen
  \bibfield  {author} {\bibinfo {author} {\bibfnamefont {P.}~\bibnamefont
  {Lodahl}}, \bibinfo {author} {\bibfnamefont {S.}~\bibnamefont {Mahmoodian}},
  \bibinfo {author} {\bibfnamefont {S.}~\bibnamefont {Stobbe}}, \bibinfo
  {author} {\bibfnamefont {A.}~\bibnamefont {Rauschenbeutel}}, \bibinfo
  {author} {\bibfnamefont {P.}~\bibnamefont {Schneeweiss}}, \bibinfo {author}
  {\bibfnamefont {J.}~\bibnamefont {Volz}}, \bibinfo {author} {\bibfnamefont
  {H.}~\bibnamefont {Pichler}},\ and\ \bibinfo {author} {\bibfnamefont
  {P.}~\bibnamefont {Zoller}},\ }\bibfield  {title} {\bibinfo {title} {Chiral
  quantum optics},\ }\href {https://doi.org/10.1038/nature21037} {\bibfield
  {journal} {\bibinfo  {journal} {Nature}\ }\textbf {\bibinfo {volume} {541}},\
  \bibinfo {pages} {473} (\bibinfo {year} {2017})}\BibitemShut {NoStop}%
\bibitem [{\citenamefont {Kariyado}\ and\ \citenamefont
  {Hu}(2017)}]{Kariyado2017}%
  \BibitemOpen
  \bibfield  {author} {\bibinfo {author} {\bibfnamefont {T.}~\bibnamefont
  {Kariyado}}\ and\ \bibinfo {author} {\bibfnamefont {X.}~\bibnamefont {Hu}},\
  }\bibfield  {title} {\bibinfo {title} {Topological states characterized by
  mirror winding numbers in graphene with bond modulation},\ }\href
  {https://doi.org/10.1038/s41598-017-16334-0} {\bibfield  {journal} {\bibinfo
  {journal} {Scientific Reports}\ }\textbf {\bibinfo {volume} {7}},\ \bibinfo
  {pages} {16515} (\bibinfo {year} {2017})}\BibitemShut {NoStop}%
\bibitem [{\citenamefont {Cherqui}\ \emph {et~al.}(2019)\citenamefont
  {Cherqui}, \citenamefont {Bourgeois}, \citenamefont {Wang},\ and\
  \citenamefont {Schatz}}]{cherqui2019plasmonic}%
  \BibitemOpen
  \bibfield  {author} {\bibinfo {author} {\bibfnamefont {C.}~\bibnamefont
  {Cherqui}}, \bibinfo {author} {\bibfnamefont {M.~R.}\ \bibnamefont
  {Bourgeois}}, \bibinfo {author} {\bibfnamefont {D.}~\bibnamefont {Wang}},\
  and\ \bibinfo {author} {\bibfnamefont {G.~C.}\ \bibnamefont {Schatz}},\
  }\bibfield  {title} {\bibinfo {title} {Plasmonic surface lattice resonances:
  Theory and computation},\ }\href
  {https://doi.org/10.1021/acs.accounts.9b00312} {\bibfield  {journal}
  {\bibinfo  {journal} {Accounts of Chemical Research}\ }\textbf {\bibinfo
  {volume} {52}},\ \bibinfo {pages} {2548} (\bibinfo {year}
  {2019})}\BibitemShut {NoStop}%
\bibitem [{\citenamefont {Yves}\ \emph {et~al.}(2020)\citenamefont {Yves},
  \citenamefont {Berthelot}, \citenamefont {Lerosey},\ and\ \citenamefont
  {Lemoult}}]{yves2020locally}%
  \BibitemOpen
  \bibfield  {author} {\bibinfo {author} {\bibfnamefont {S.}~\bibnamefont
  {Yves}}, \bibinfo {author} {\bibfnamefont {T.}~\bibnamefont {Berthelot}},
  \bibinfo {author} {\bibfnamefont {G.}~\bibnamefont {Lerosey}},\ and\ \bibinfo
  {author} {\bibfnamefont {F.}~\bibnamefont {Lemoult}},\ }\bibfield  {title}
  {\bibinfo {title} {Locally polarized wave propagation through crystalline
  metamaterials},\ }\href {https://doi.org/10.1103/PhysRevB.101.035127}
  {\bibfield  {journal} {\bibinfo  {journal} {Phys. Rev. B}\ }\textbf {\bibinfo
  {volume} {101}},\ \bibinfo {pages} {035127} (\bibinfo {year}
  {2020})}\BibitemShut {NoStop}%
\bibitem [{\citenamefont {Baranov}\ \emph {et~al.}(2017)\citenamefont
  {Baranov}, \citenamefont {Savelev}, \citenamefont {Li}, \citenamefont
  {Krasnok},\ and\ \citenamefont {Alù}}]{baranov2017modifying}%
  \BibitemOpen
  \bibfield  {author} {\bibinfo {author} {\bibfnamefont {D.~G.}\ \bibnamefont
  {Baranov}}, \bibinfo {author} {\bibfnamefont {R.~S.}\ \bibnamefont
  {Savelev}}, \bibinfo {author} {\bibfnamefont {S.~V.}\ \bibnamefont {Li}},
  \bibinfo {author} {\bibfnamefont {A.~E.}\ \bibnamefont {Krasnok}},\ and\
  \bibinfo {author} {\bibfnamefont {A.}~\bibnamefont {Alù}},\ }\bibfield
  {title} {\bibinfo {title} {Modifying magnetic dipole spontaneous emission
  with nanophotonic structures},\ }\href
  {https://doi.org/10.1002/lpor.201600268} {\bibfield  {journal} {\bibinfo
  {journal} {Laser \& Photonics Reviews}\ }\textbf {\bibinfo {volume} {11}},\
  \bibinfo {pages} {1600268} (\bibinfo {year} {2017})}\BibitemShut {NoStop}%
\bibitem [{\citenamefont {Alaee}\ \emph {et~al.}(2020)\citenamefont {Alaee},
  \citenamefont {Gurlek}, \citenamefont {Albooyeh}, \citenamefont
  {Mart\'{\i}n-Cano},\ and\ \citenamefont {Sandoghdar}}]{alaee2020quantum}%
  \BibitemOpen
  \bibfield  {author} {\bibinfo {author} {\bibfnamefont {R.}~\bibnamefont
  {Alaee}}, \bibinfo {author} {\bibfnamefont {B.}~\bibnamefont {Gurlek}},
  \bibinfo {author} {\bibfnamefont {M.}~\bibnamefont {Albooyeh}}, \bibinfo
  {author} {\bibfnamefont {D.}~\bibnamefont {Mart\'{\i}n-Cano}},\ and\ \bibinfo
  {author} {\bibfnamefont {V.}~\bibnamefont {Sandoghdar}},\ }\bibfield  {title}
  {\bibinfo {title} {Quantum metamaterials with magnetic response at optical
  frequencies},\ }\href {https://doi.org/10.1103/PhysRevLett.125.063601}
  {\bibfield  {journal} {\bibinfo  {journal} {Phys. Rev. Lett.}\ }\textbf
  {\bibinfo {volume} {125}},\ \bibinfo {pages} {063601} (\bibinfo {year}
  {2020})}\BibitemShut {NoStop}%
\bibitem [{\citenamefont {Garc\'{i}a-Etxarri}\ \emph
  {et~al.}(2011)\citenamefont {Garc\'{i}a-Etxarri}, \citenamefont
  {G\'{o}mez-Medina}, \citenamefont {Froufe-P\'{e}rez}, \citenamefont
  {L\'{o}pez}, \citenamefont {Chantada}, \citenamefont {Scheffold},
  \citenamefont {Aizpurua}, \citenamefont {Nieto-Vesperinas},\ and\
  \citenamefont {S\'{a}enz}}]{garcia2011strong}%
  \BibitemOpen
  \bibfield  {author} {\bibinfo {author} {\bibfnamefont {A.}~\bibnamefont
  {Garc\'{i}a-Etxarri}}, \bibinfo {author} {\bibfnamefont {R.}~\bibnamefont
  {G\'{o}mez-Medina}}, \bibinfo {author} {\bibfnamefont {L.~S.}\ \bibnamefont
  {Froufe-P\'{e}rez}}, \bibinfo {author} {\bibfnamefont {C.}~\bibnamefont
  {L\'{o}pez}}, \bibinfo {author} {\bibfnamefont {L.}~\bibnamefont {Chantada}},
  \bibinfo {author} {\bibfnamefont {F.}~\bibnamefont {Scheffold}}, \bibinfo
  {author} {\bibfnamefont {J.}~\bibnamefont {Aizpurua}}, \bibinfo {author}
  {\bibfnamefont {M.}~\bibnamefont {Nieto-Vesperinas}},\ and\ \bibinfo {author}
  {\bibfnamefont {J.~J.}\ \bibnamefont {S\'{a}enz}},\ }\bibfield  {title}
  {\bibinfo {title} {Strong magnetic response of submicron silicon particles in
  the infrared},\ }\href {https://doi.org/10.1364/OE.19.004815} {\bibfield
  {journal} {\bibinfo  {journal} {Opt. Express}\ }\textbf {\bibinfo {volume}
  {19}},\ \bibinfo {pages} {4815} (\bibinfo {year} {2011})}\BibitemShut
  {NoStop}%
\bibitem [{\citenamefont {Kim}\ and\ \citenamefont
  {Rho}(2020)}]{minkyung2020quantum}%
  \BibitemOpen
  \bibfield  {author} {\bibinfo {author} {\bibfnamefont {M.}~\bibnamefont
  {Kim}}\ and\ \bibinfo {author} {\bibfnamefont {J.}~\bibnamefont {Rho}},\
  }\bibfield  {title} {\bibinfo {title} {Quantum hall phase and chiral edge
  states simulated by a coupled dipole method},\ }\href
  {https://doi.org/10.1103/PhysRevB.101.195105} {\bibfield  {journal} {\bibinfo
   {journal} {Phys. Rev. B}\ }\textbf {\bibinfo {volume} {101}},\ \bibinfo
  {pages} {195105} (\bibinfo {year} {2020})}\BibitemShut {NoStop}%
\bibitem [{\citenamefont {Deng}\ \emph {et~al.}(2017)\citenamefont {Deng},
  \citenamefont {Chen}, \citenamefont {Zhao},\ and\ \citenamefont
  {Dong}}]{deng2017transverse}%
  \BibitemOpen
  \bibfield  {author} {\bibinfo {author} {\bibfnamefont {W.-M.}\ \bibnamefont
  {Deng}}, \bibinfo {author} {\bibfnamefont {X.-D.}\ \bibnamefont {Chen}},
  \bibinfo {author} {\bibfnamefont {F.-L.}\ \bibnamefont {Zhao}},\ and\
  \bibinfo {author} {\bibfnamefont {J.-W.}\ \bibnamefont {Dong}},\ }\bibfield
  {title} {\bibinfo {title} {Transverse angular momentum in topological
  photonic crystals},\ }\href {https://doi.org/10.1088/2040-8986/aa9b06}
  {\bibfield  {journal} {\bibinfo  {journal} {Journal of Optics}\ }\textbf
  {\bibinfo {volume} {20}},\ \bibinfo {pages} {014006} (\bibinfo {year}
  {2017})}\BibitemShut {NoStop}%
\bibitem [{\citenamefont {Deng}\ \emph {et~al.}(2019)\citenamefont {Deng},
  \citenamefont {Chen}, \citenamefont {Chen}, \citenamefont {Zhao},\ and\
  \citenamefont {Dong}}]{deng2019vortex}%
  \BibitemOpen
  \bibfield  {author} {\bibinfo {author} {\bibfnamefont {W.-M.}\ \bibnamefont
  {Deng}}, \bibinfo {author} {\bibfnamefont {X.-D.}\ \bibnamefont {Chen}},
  \bibinfo {author} {\bibfnamefont {W.-J.}\ \bibnamefont {Chen}}, \bibinfo
  {author} {\bibfnamefont {F.-L.}\ \bibnamefont {Zhao}},\ and\ \bibinfo
  {author} {\bibfnamefont {J.-W.}\ \bibnamefont {Dong}},\ }\bibfield  {title}
  {\bibinfo {title} {Vortex index identification and unidirectional propagation
  in kagome photonic crystals},\ }\href
  {https://doi.org/10.1515/nanoph-2019-0009} {\bibfield  {journal} {\bibinfo
  {journal} {Nanophotonics}\ }\textbf {\bibinfo {volume} {8}},\ \bibinfo
  {pages} {833 } (\bibinfo {year} {2019})}\BibitemShut {NoStop}%
\bibitem [{\citenamefont {Chen}\ \emph {et~al.}(2017)\citenamefont {Chen},
  \citenamefont {Zhao}, \citenamefont {Chen},\ and\ \citenamefont
  {Dong}}]{chen2017valley}%
  \BibitemOpen
  \bibfield  {author} {\bibinfo {author} {\bibfnamefont {X.-D.}\ \bibnamefont
  {Chen}}, \bibinfo {author} {\bibfnamefont {F.-L.}\ \bibnamefont {Zhao}},
  \bibinfo {author} {\bibfnamefont {M.}~\bibnamefont {Chen}},\ and\ \bibinfo
  {author} {\bibfnamefont {J.-W.}\ \bibnamefont {Dong}},\ }\bibfield  {title}
  {\bibinfo {title} {Valley-contrasting physics in all-dielectric photonic
  crystals: Orbital angular momentum and topological propagation},\ }\href
  {https://doi.org/10.1103/PhysRevB.96.020202} {\bibfield  {journal} {\bibinfo
  {journal} {Phys. Rev. B}\ }\textbf {\bibinfo {volume} {96}},\ \bibinfo
  {pages} {020202} (\bibinfo {year} {2017})}\BibitemShut {NoStop}%
\bibitem [{\citenamefont {Chen}\ \emph {et~al.}(2018)\citenamefont {Chen},
  \citenamefont {Shi}, \citenamefont {Liu}, \citenamefont {Lu}, \citenamefont
  {Deng}, \citenamefont {Dai}, \citenamefont {Cheng},\ and\ \citenamefont
  {Dong}}]{chen2018tunable}%
  \BibitemOpen
  \bibfield  {author} {\bibinfo {author} {\bibfnamefont {X.-D.}\ \bibnamefont
  {Chen}}, \bibinfo {author} {\bibfnamefont {F.-L.}\ \bibnamefont {Shi}},
  \bibinfo {author} {\bibfnamefont {H.}~\bibnamefont {Liu}}, \bibinfo {author}
  {\bibfnamefont {J.-C.}\ \bibnamefont {Lu}}, \bibinfo {author} {\bibfnamefont
  {W.-M.}\ \bibnamefont {Deng}}, \bibinfo {author} {\bibfnamefont {J.-Y.}\
  \bibnamefont {Dai}}, \bibinfo {author} {\bibfnamefont {Q.}~\bibnamefont
  {Cheng}},\ and\ \bibinfo {author} {\bibfnamefont {J.-W.}\ \bibnamefont
  {Dong}},\ }\bibfield  {title} {\bibinfo {title} {Tunable electromagnetic flow
  control in valley photonic crystal waveguides},\ }\href
  {https://doi.org/10.1103/PhysRevApplied.10.044002} {\bibfield  {journal}
  {\bibinfo  {journal} {Phys. Rev. Applied}\ }\textbf {\bibinfo {volume}
  {10}},\ \bibinfo {pages} {044002} (\bibinfo {year} {2018})}\BibitemShut
  {NoStop}%
\bibitem [{\citenamefont {Ye}\ \emph {et~al.}(2017)\citenamefont {Ye},
  \citenamefont {Yang}, \citenamefont {Hong~Hang}, \citenamefont {Qiu},\ and\
  \citenamefont {Liu}}]{ye2017observation}%
  \BibitemOpen
  \bibfield  {author} {\bibinfo {author} {\bibfnamefont {L.}~\bibnamefont
  {Ye}}, \bibinfo {author} {\bibfnamefont {Y.}~\bibnamefont {Yang}}, \bibinfo
  {author} {\bibfnamefont {Z.}~\bibnamefont {Hong~Hang}}, \bibinfo {author}
  {\bibfnamefont {C.}~\bibnamefont {Qiu}},\ and\ \bibinfo {author}
  {\bibfnamefont {Z.}~\bibnamefont {Liu}},\ }\bibfield  {title} {\bibinfo
  {title} {Observation of valley-selective microwave transport in photonic
  crystals},\ }\href {https://doi.org/10.1063/1.5009597} {\bibfield  {journal}
  {\bibinfo  {journal} {Applied Physics Letters}\ }\textbf {\bibinfo {volume}
  {111}},\ \bibinfo {pages} {251107} (\bibinfo {year} {2017})}\BibitemShut
  {NoStop}%
\bibitem [{\citenamefont {Chervy}\ \emph {et~al.}(2018)\citenamefont {Chervy},
  \citenamefont {Azzini}, \citenamefont {Lorchat}, \citenamefont {Wang},
  \citenamefont {Gorodetski}, \citenamefont {Hutchison}, \citenamefont
  {Berciaud}, \citenamefont {Ebbesen},\ and\ \citenamefont
  {Genet}}]{chervy2018room}%
  \BibitemOpen
  \bibfield  {author} {\bibinfo {author} {\bibfnamefont {T.}~\bibnamefont
  {Chervy}}, \bibinfo {author} {\bibfnamefont {S.}~\bibnamefont {Azzini}},
  \bibinfo {author} {\bibfnamefont {E.}~\bibnamefont {Lorchat}}, \bibinfo
  {author} {\bibfnamefont {S.}~\bibnamefont {Wang}}, \bibinfo {author}
  {\bibfnamefont {Y.}~\bibnamefont {Gorodetski}}, \bibinfo {author}
  {\bibfnamefont {J.~A.}\ \bibnamefont {Hutchison}}, \bibinfo {author}
  {\bibfnamefont {S.}~\bibnamefont {Berciaud}}, \bibinfo {author}
  {\bibfnamefont {T.~W.}\ \bibnamefont {Ebbesen}},\ and\ \bibinfo {author}
  {\bibfnamefont {C.}~\bibnamefont {Genet}},\ }\bibfield  {title} {\bibinfo
  {title} {Room temperature chiral coupling of valley excitons with
  spin-momentum locked surface plasmons},\ }\href
  {https://doi.org/10.1021/acsphotonics.7b01032} {\bibfield  {journal}
  {\bibinfo  {journal} {ACS Photonics}\ }\textbf {\bibinfo {volume} {5}},\
  \bibinfo {pages} {1281} (\bibinfo {year} {2018})}\BibitemShut {NoStop}%
\bibitem [{\citenamefont {Hu}\ \emph {et~al.}(2019)\citenamefont {Hu},
  \citenamefont {Hong}, \citenamefont {Wang}, \citenamefont {Wu}, \citenamefont
  {Xu}, \citenamefont {Zhao}, \citenamefont {Liu}, \citenamefont {Zhang},
  \citenamefont {Garcia-Vidal}, \citenamefont {Wang}, \citenamefont {Lu},\ and\
  \citenamefont {Qiu}}]{hu2019coherent}%
  \BibitemOpen
  \bibfield  {author} {\bibinfo {author} {\bibfnamefont {G.}~\bibnamefont
  {Hu}}, \bibinfo {author} {\bibfnamefont {X.}~\bibnamefont {Hong}}, \bibinfo
  {author} {\bibfnamefont {K.}~\bibnamefont {Wang}}, \bibinfo {author}
  {\bibfnamefont {J.}~\bibnamefont {Wu}}, \bibinfo {author} {\bibfnamefont
  {H.-X.}\ \bibnamefont {Xu}}, \bibinfo {author} {\bibfnamefont
  {W.}~\bibnamefont {Zhao}}, \bibinfo {author} {\bibfnamefont {W.}~\bibnamefont
  {Liu}}, \bibinfo {author} {\bibfnamefont {S.}~\bibnamefont {Zhang}}, \bibinfo
  {author} {\bibfnamefont {F.}~\bibnamefont {Garcia-Vidal}}, \bibinfo {author}
  {\bibfnamefont {B.}~\bibnamefont {Wang}}, \bibinfo {author} {\bibfnamefont
  {P.}~\bibnamefont {Lu}},\ and\ \bibinfo {author} {\bibfnamefont {C.-W.}\
  \bibnamefont {Qiu}},\ }\bibfield  {title} {\bibinfo {title} {Coherent
  steering of nonlinear chiral valley photons with a synthetic au--ws2
  metasurface},\ }\href {https://doi.org/10.1038/s41566-019-0399-1} {\bibfield
  {journal} {\bibinfo  {journal} {Nature Photonics}\ }\textbf {\bibinfo
  {volume} {13}},\ \bibinfo {pages} {467} (\bibinfo {year} {2019})}\BibitemShut
  {NoStop}%
\bibitem [{\citenamefont {Sun}\ \emph {et~al.}(2019)\citenamefont {Sun},
  \citenamefont {Wang}, \citenamefont {Krasnok}, \citenamefont {Choi},
  \citenamefont {Shi}, \citenamefont {Gomez-Diaz}, \citenamefont {Zepeda},
  \citenamefont {Gwo}, \citenamefont {Shih}, \citenamefont {Al{\`u}},\ and\
  \citenamefont {Li}}]{sun2019separation}%
  \BibitemOpen
  \bibfield  {author} {\bibinfo {author} {\bibfnamefont {L.}~\bibnamefont
  {Sun}}, \bibinfo {author} {\bibfnamefont {C.-Y.}\ \bibnamefont {Wang}},
  \bibinfo {author} {\bibfnamefont {A.}~\bibnamefont {Krasnok}}, \bibinfo
  {author} {\bibfnamefont {J.}~\bibnamefont {Choi}}, \bibinfo {author}
  {\bibfnamefont {J.}~\bibnamefont {Shi}}, \bibinfo {author} {\bibfnamefont
  {J.~S.}\ \bibnamefont {Gomez-Diaz}}, \bibinfo {author} {\bibfnamefont
  {A.}~\bibnamefont {Zepeda}}, \bibinfo {author} {\bibfnamefont
  {S.}~\bibnamefont {Gwo}}, \bibinfo {author} {\bibfnamefont {C.-K.}\
  \bibnamefont {Shih}}, \bibinfo {author} {\bibfnamefont {A.}~\bibnamefont
  {Al{\`u}}},\ and\ \bibinfo {author} {\bibfnamefont {X.}~\bibnamefont {Li}},\
  }\bibfield  {title} {\bibinfo {title} {Separation of valley excitons in a
  mos2 monolayer using a subwavelength asymmetric groove array},\ }\href
  {https://doi.org/10.1038/s41566-019-0348-z} {\bibfield  {journal} {\bibinfo
  {journal} {Nature Photonics}\ }\textbf {\bibinfo {volume} {13}},\ \bibinfo
  {pages} {180} (\bibinfo {year} {2019})}\BibitemShut {NoStop}%
\bibitem [{\citenamefont {{COMSOL AB, Stockholm}}()}]{comsol}%
  \BibitemOpen
  \bibfield  {author} {\bibinfo {author} {\bibnamefont {{COMSOL AB,
  Stockholm}}},\ }\href {www.comsol.com} {\bibinfo {title} {{RF Module - COMSOL
  Multiphysics}}}\BibitemShut {NoStop}%
\end{thebibliography}%

\end{document}